\documentclass[aps,prd,twocolumn, 
  superscriptaddress]{revtex4}
\usepackage{float}
\usepackage{multirow}
\usepackage{diagbox}
\usepackage{graphicx} \usepackage{color} \usepackage{natbib}
\usepackage{epsfig}
\usepackage{amsmath,amssymb,amsfonts}
\def\prl{Phys. Rev. Lett.}
\def\prd{Phys. Rev. D}
\def\cqg{Class. Quantum Grav.}

\begin{document}

\title{Aspherical deformations of the Choptuik spacetime}

\author{Thomas W. Baumgarte}
\affiliation{Department of Physics and Astronomy, Bowdoin College, Brunswick, ME 04011, USA}

\begin{abstract}
We perform dynamical and nonlinear numerical simulations to study critical phenomena in the gravitational collapse of massless scalar fields in the absence of spherical symmetry.  We evolve axisymmetric sets of initial data and examine the effects of deviation from spherical symmetry.  For small deviations we find values for the critical exponent and echoing period of the discretely self-similar critical solution that agree well with established values; moreover we find that such small deformations behave like damped oscillations whose damping coefficient and oscillation frequencies are consistent with those predicted in the linear perturbation calculations of Mart\'in-Garc\'ia and Gundlach.  However, we also find that the critical exponent and echoing period appear to decrease with increasing departure from sphericity, and that, for sufficiently large departures from spherical symmetry, the deviations become unstable and grow, confirming earlier results by Choptuik {\it et.al.}.   We find some evidence that these growing modes lead to a bifurcation, similar to those reported by Choptuik {\it et.al.}, with two centers of collapse forming on the symmetry axis above and below the origin.  These findings suggest that nonlinear perturbations of the critical solution lead to changes in the effective values of the critical exponent, echoing period and damping coefficient, and may even change the sign of the latter, so that perturbations that are stable in the linear regime can become unstable in the nonlinear regime.
\end{abstract}


\maketitle

\section{Introduction}
\label{sec:intro}

Critical phenomena in gravitational collapse were first reported in Choptuik's seminal study of massless scalar fields \cite{Cho93}.  Specifically, Choptuik performed numerical simulations of the dynamical evolution of massless scalar fields coupled to Einstein's equations.  He considered different families of spherically symmetric initial data, parameterized by a parameter $\eta$, say.  Sufficiently strong initial data collapse to form a black hole, while sufficiently weak data disperse to infinity and leave behind flat space.   We refer to a value of the parameter $\eta$ that separates supercritical from subcritical data as a critical parameter $\eta_*$.

Critical phenomena appear close to the black-hole threshold, in the vicinity of a critical parameter $\eta_*$ (see, e.g., \cite{Gun03,GunM07} for reviews).  For many different systems and matter models, and with $\eta$ close to $\eta_*$, the dynamical evolution will, during an intermediate time, approximately follow a critical solution that contracts in a self-similar fashion.  This self-similar solution focusses on an accumulation event that occurs at a finite proper time $\tau_*$ as measured by an observer at the center.   At any proper time $\tau$, the spatial scale $R_{\rm sc}$ of the critical solution is therefore proportional to $\tau_* - \tau$,
\begin{equation} \label{selfsim}
R_{\rm sc} \simeq \tau_* - \tau
\end{equation}
(here and throughout we assume geometrized units, in which $c = 1$ and $G = 1$).   For massless scalar fields we refer to this self-similar critical solution as the Choptuik spacetime.  The better the fine-tuning to criticality, i.e.~the closer $\eta$ to $\eta_*$, the longer the evolution will follow the critical solution, and the later the time $\tau$ at which the evolution will depart from the critical solution.  The scale $R_{\rm sc} \simeq \tau_* - \tau$ of the critical solution at this time $\tau$ imprints a length scale on the subsequent evolution, resulting, for example, in the famous scaling law
\begin{equation} \label{mass_scaling}
M \simeq |\eta - \eta_*|^\gamma
\end{equation}
for the black-hole mass $M$ in supercritical evolutions.  Here $\gamma$ is the critical exponent, which Choptuik found to be about $\gamma \simeq 0.37$ in his original numerical experiments, independently of the parametrization of the initial data.

The critical exponent can also be found by considering perturbations of the critical solution.  In spherical symmetry, the departure from the critical solution is caused by an unstable, spherically symmetric mode that grows exponentially with $\exp(\kappa T)$, 
where
\begin{equation} \label{T}
T = - \ln( \tau_* - \tau).
\end{equation}
Note that $\tau \rightarrow \tau_*$ corresponds to $T \rightarrow \infty$.  The critical exponent can then be shown to be the inverse of the Lyapunov exponent $\kappa$, i.e.~$\gamma = 1/\kappa$.  For a massless scalar field, Gundlach found $\gamma = 0.374(1)$ from such a perturbative calculation \cite{Gun97}, in excellent agreement with Choptuik's numerical results.

Choptuik's discovery launched an entire new field of research, triggering a large number of studies, both numerical and perturbative, of critical phenomena for different matter models, spacetime dimensions and asymptotics (see \cite{Gun03,GunM07} for reviews).  In particular, these studies showed that for some matter models, for example radiation fluids \cite{EvaC94}, the critical solution displays a continuous self-similarity, while for others, including massless scalar fields, a discrete self-similarity.   A discretely self-similar solution performs an oscillation as it contracts towards the accumulation event.  In terms of the logarithmic time $T$, this oscillation has a period $\Delta$, which can be determined by casting the problem as an eigenvalue problem \cite{Gun95,Gun97}.  The value computed by \cite{MarG03} is $\Delta = 3.445452402(3)$.  At a time $T + \Delta$, the self-similar solution will take the same shape as at time $T$, but on a radial scale that is smaller than that at $T$ by a factor of 
\begin{eqnarray} \label{R_echoes}
\frac{R_{\rm sc}(T + \Delta)}{R_{\rm sc}(T)} & = & \frac{\tau_* - \tau(T + \Delta)}{\tau_* - \tau(T)} = \frac{e^{-(T + \Delta)}}{e^{-T}} \nonumber \\
 & = & e^{- \Delta} \sim \frac{1}{31.3}.
\end{eqnarray}
The reappearance of the solution at time intervals $\Delta$ is referred to as echoing, and $\Delta$ as the echoing period.  

Until recently most numerical studies of critical collapse assumed spherical symmetry, which is helpful for resolving the small spatial structures that emerge close to criticality (a notable early exception is the study by Abrahams and Evans of critical phenomena in the collapse of gravitational waves \cite{AbrE93}).   A number of important questions, however, cannot even be addressed under the assumptions of spherical symmetry.  One such question concerns the stability of the critical solution to aspherical modes.  

Adopting a perturbative treatment, Mart\'in-Garc\'ia and Gundlach \cite{MarG99} (hereafter MGG) found that all aspherical modes in the collapse of massless scalar fields are stable, leading them to conclude that ``all nonspherical perturbations of the Choptuik spacetime decay".  We note, however, that the decay rate of the most slowly damped mode, an $\ell = 2$ mode, was found to be quite small in magnitude, corresponding to a slow damping.

In \cite{ChoHLP03b}, Choptuik {\it et.al.} (hereafter CHLP) adopted a code in cylindrical coordinates (see also \cite{ChoHLP03a}) to study critical collapse of massless scalar fields in axisymmetry.  CHLP adopted a two-parameter family of initial data, with $\eta$ parametrizing the overall strength of the data, and $\epsilon$ the departure from spherical symmetry (see Eq.~\ref{indata} below).   For a given value of $\epsilon$, CHLP then fine-tuned $\eta$ to the black-hole threshold (to about $\eta/\eta_* - 1 \simeq 10^{-15}$ for small values of $\epsilon$) and studied the properties of the emerging critical solution.  

Briefly summarized, CHLP reported two key results.  One of their findings is that the critical exponent $\gamma$ and the period $\Delta$ appear to depend on $\epsilon$, with both of them decreasing with increasing departure from sphericity (see Table I in CHLP).  The other, perhaps more important finding is that, for large values of $\epsilon$ and for exquisite fine-tuning to criticality, the collapsing region in the center of the spacetime appears to bifurcate into two collapsing regions, located along the axis of symmetry.  This result indicates that there exists a non-spherical growing mode that dominates the evolution at late times $T$, in apparent conflict with the findings of MGG.  

Given the seemingly contradictory results of MGG and CHLP it is of interest to verify whether the results of CHLP can be reproduced with an independent numerical code.  However, only few codes have been able to simulate critical collapse in the absence of spherical symmetry, as most recent numerical relativity codes in three spatial dimensions have been designed for simulations of binary problems (but see \cite{AlcABLSST00,HilBWDBMM13} for attempts with such codes in the context of vacuum evolutions, as well as \cite{Sor10,Sor11,HilWB15,HilWB17}  for examples of recent codes specifically designed for simulations of critical collapse in vacuum spacetimes).  In \cite{HeaL14}, Healy and Laguna simulated critical collapse of scalar fields in three spatial dimensions (see also \cite{CloL16}).   Their calculations achieved more modest fine-tuning (to about $\eta/\eta_* - 1 \simeq 10^{-4}$), which was sufficient to measure the critical exponent $\gamma$ and possibly the echoing period $\Delta$, but not to make statements about the stability of the critical solution.  More recently, Deppe {\it et.al.}\ \cite{DepKST18} performed similar simulations of critical collapse of scalar fields in three spatial dimensions, and achieved better fine-tuning (to about $\eta/\eta_* - 1 \simeq 10^{-6}$).  Interestingly, their reported values for the echoing period differ slightly from those found in previous numerical solutions as well as the semi-analytical value of \cite{MarG03}.  Deppe {\it et.al.}\ did not observe any growing modes, but it also is not clear whether their fine-tuning would be sufficient to find such modes.  

In this paper we adopt a code in spherical polar coordinates \cite{BauMCM13,BauMM15} to study critical collapse of massless scalar fields in the absence of spherical symmetry.  The code has been used previously for studies of critical collapse in ultrarelativistic fluids \cite{BauM15,BauG16,GunB16,GunB18}.  In essence, we follow a suggestion made in the Conclusion section of CHLP, namely ``... one could write a code adapted to the spherical critical solution (for instance using spherical polar coordinates with a logarithmic radial coordinate)."  We evolve the same axisymmetric initial data as those considered by CHLP (see Eq.~\ref{indata} below), and study the properties of near-critical solutions, fine-tuned to about $\eta/\eta_* - 1 \simeq 10^{-12}$ or better, for different deviations $\epsilon$ from sphericity.    

For small values of $\epsilon$ we find values of the critical exponent $\gamma$ and the echoing period $\Delta$ that agree well with those found in most previous studies; furthermore we find that small deformations from sphericity behave as damped oscillations with a decay rate $\kappa$ and oscillation frequencies similar to those computed by MGG.  We also confirm CHLP's finding that both $\gamma$ and $\Delta$ appear to decrease with increasing $\epsilon$.  For sufficiently large values of $\epsilon$ we find evidence for a growing unstable mode, in further agreement with CHLP's results.   We also present some evidence that this growing mode results in a bifurcation similar to that reported by CHLP, with two centers of oscillation forming on the symmetry axis above and below the origin.   These results suggest that, in the presence of nonlinear perturbations of the spherically symmetric critical solution, the evolution can be described in terms of effective values of $\gamma$, $\Delta$ and $\kappa$ that depend on the departure from spherical symmetry.  The decay rate $\kappa$, which started out quite close to zero, may even change sign, thereby making a mode that was stable in the linear regime, as predicted by MGG, unstable for sufficiently large departures from spherical symmetry, as observed by CHLP.


This paper is organized as follows.  We present basic equations, our numerical method, initial data and diagnostics in Section \ref{sec:setup}. In Section \ref{sec:results} we then present the results from our simulations, first under the assumption of spherical symmetry and then relaxing this assumption.  We briefly summarize in Section \ref{sec:summary}.  

\section{Setup of the problem}
\label{sec:setup}

\subsection{Basic equations}
\label{sec:basics}

We solve Einstein's equations 
\begin{equation} \label{einstein}
G_{ab} = 8 \pi T_{ab},
\end{equation}
where $G_{ab}$ is the Einstein tensor, under the assumption that the stress-energy tensor $T_{ab}$ is that of a massless scalar field (see Eq.~(\ref{stressenergy}) below).  Specifically, we adopt the Baumgarte-Shapiro-Shibata-Nakamura (BSSN) formulation of Einstein's equations \cite{NakOK87,ShiN95,BauS98} in spherical polar coordinates $(r,\theta,\varphi)$ with the help of a reference-metric formulation \cite{BonGGN04,ShiUF04,Bro09,Gou12}.  The BSSN formalism is based on a ``3+1" decomposition of the spacetime, so that the line element takes the form
\begin{equation}
ds^2 = g_{ab} dx^a dx^b = - \alpha^2 dt^2 + \gamma_{ij} (dx^i + \beta^i dt)(dx^j + \beta^j dt).
\end{equation}
Here $g_{ab}$ is the spacetime metric, $\alpha$ the lapse function, $\gamma_{ij}$ the spatial metric, and $\beta^i$ the shift vector.   The normal on the spatial slices is given by
\begin{equation}
n_a = (-\alpha,0,0,0) \mbox{~~~or~~~} n^a = \alpha^{-1} (1, - \beta^i).
\end{equation}
We further introduce a conformal decomposition of the spatial metric
\begin{equation}
\gamma_{ij} = \psi^4 \bar \gamma_{ij},
\end{equation}
where $\psi$ is the conformal factor and $\bar \gamma_{ij}$ the conformally related metric.  For our applications in spherical coordinates we also identify the reference metric $\hat \gamma_{ij}$ with the flat metric $\eta_{ij}$ expressed in spherical coordinates.  We refer the reader to \cite{BauMCM13,BauMM15} for details of this formalism, as well as its implementation in our code.

The massless scalar field $\phi$ satisfies the wave equation
\begin{equation} \label{wave}
g^{ab} \nabla_a \nabla_b \phi = 0,
\end{equation}
where $\nabla_a$ is the covariant derivative associated with the spacetime metric $g_{ab}$, and its
stress-energy tensor is given by
\begin{equation} \label{stressenergy}
T_{ab} = (\nabla_a \phi)(\nabla_b \phi) - \frac{1}{2} g_{ab} (\nabla_c \phi)(\nabla^c \phi).
\end{equation}
Our convention for $T_{ab}$ agrees with that used in many places in the literature, including \cite{HawE73} (as well as, for example, \cite{AkbC15,DepKST18}), but differs from that of CHLP by a factor of 2 (see their Eq.~3).   In comparisons of our results with theirs the scalar field $\phi$ therefore needs to be rescaled with a factor of $\sqrt{2}$.  

\subsection{Initial data}
\label{sec:indata}

We impose initial data at an initial moment of time symmetry and assume that, at this time, the scalar field is given by the axisymmetric two-parameter family
\begin{equation} \label{indata}
\phi = \eta \exp\left(- (r / r_0)^2 (\sin^2 \theta + (1 - \epsilon^2) \cos^2 \theta) \right)
\end{equation}
(compare Eq.~9 in CHLP).  The parameter $\eta$ determines the overall strength of the initial data, and $\epsilon^2$ the departure from sphericity.   The constant $r_0$ has units of length; in practice we set $r_0 = 1$, meaning that we report all dimensional results in units of $r_0$.  


We also assume the spatial metric to be conformally flat initially, $\bar \gamma_{ij} = \hat \gamma_{ij}$, and, consistent with time symmetry, choose the shift vector $\beta^i$ as well as time derivatives of the spatial metric and the scalar field to vanish.  The initial conformal factor $\psi$ can then be found by solving the Hamiltonian constraint
\begin{equation} \label{Hamiltonian}
\hat D^2 \psi = - 2 \pi \psi^5 \rho
\end{equation}
subject to Robin boundary conditions at the outer boundary.  Here $\hat D^2$ is the flat Laplace operator associated with $\hat \gamma_{ij}$, and $\rho$ is the energy density
\begin{equation} \label{rho}
\rho = n_a n_b T^{ab}.
\end{equation}
We use an iterative process, alternately solving Eq.~(\ref{Hamiltonian}) and updating (\ref{rho}), to construct simultaneous solutions to the two equations.


We complete the specification of the initial data by choosing a ``pre-collapsed lapse" $\alpha = \psi^{-2}$ as the initial data for the lapse function $\alpha$, as well as zero shift, $\beta^i = 0$.  

\subsection{Numerical Evolution}
\label{sec:numerics}

Our numerical code is very similar to that used in \cite{BauG16,GunB16,GunB18,CelB18}, except that here we couple Einstein's equations to a massless scalar field instead of a fluid.  We evolve the gravitational fields using the code described in \cite{BauMCM13,BauMM15}.  The code uses finite-difference methods to solve the BSSN equations \cite{NakOK87,ShiN95,BauS98}, recast with the help of a reference-metric formulation \cite{BonGGN04,ShiUF04,Bro09,Gou12}, in spherical coordinates.  We also rescale all tensorial variables, so that the coordinate singularities at the origin and on the axis can be handled analytically.  

We define a new auxiliary variable 
\begin{equation}
\Pi \equiv - n^a \nabla_a \phi = - \frac{1}{\alpha} (\partial_t \phi - \beta^i \nabla_i \phi )
\end{equation}
in order to cast the wave equation (\ref{wave}) as a pair of two equations that are first order in time.  Unlike in many other applications, however, we do not introduce variables that absorb first spatial derivatives of the scalar field $\phi$, and instead leave the spatial derivatives in (\ref{wave}) in terms of second derivatives -- similar to the treatment of the gravitational fields in the BSSN equations.

The code does not make any symmetry assumptions, but for the axisymmetric simulations presented in this paper we set to zero all derivatives with respect to $\varphi$ and use only a single grid point in the azimuthal direction.   We also impose an equatorial symmetry, and evolve only one of the two hemispheres.  For spherically symmetric simulations we use the minimum number of grid points possible in the $\theta$-direction, $N_\theta = 2$, while in the absence of spherical symmetry we use $N_\theta = 12$ uniformly allocated angular grid points for $\epsilon^2 = 0.01$ and 0.5, and $N_\theta = 14$ for $\epsilon^2 = 0.75$ (unless noted otherwise).

The radial grid points are allocated logarithmically (see Appendix A in \cite{BauM15}), so that each each grid cell is larger than its inner neighbor by a factor of $c = 1.025$.  For most simulations we use $N_r = 312$ grid points, and initially place the outer boundary at $r_{\rm out}^{\rm init} = 64$ (in our code units).   We allow the code to regrid whenever the length-scale $l = (\phi_{,rr})^{-1/2}$ becomes smaller than $25 \Delta r$, where $\Delta r$ is the innermost grid size.  In each regridding, the code variables are interpolated to a new grid that extends to a smaller outer boundary $r_{\rm out}$.  We allow up to 20 regrids, shrinking the outer boundary by equal factors down to $r_{\rm out}^{\rm final} = 0.32$.  We terminate all runs before the center of the simulation comes into causal contact with the outer boundary.   For larger $\epsilon^2$ the accumulation event occurs at a later proper time $\tau_*$ (see Table \ref{table:results} below), and we therefore placed the outer boundary at $r_{\rm out}^{\rm init} = 128$, used $N_r = 340$ radial grid points (in order to achieve the same resolution at the center), and allowed 25 regrids to the same final outer boundary location of $r_{\rm out}^{\rm final} = 0.32$.  

For this latter setup, the ratio between the radial grid size at the outer boundary and that at the origin is $c^{N_r} \simeq 4430$.  Further taking into account the regridding, we find that the ratio between initial grid size at the outer boundary and the final grid size at the origin is $ (r_{\rm out}^{\rm init} / r_{\rm out}^{\rm final} ) \, c^{N_r} \simeq 1.77 \times 10^6$.   For comparison, a code with adaptive-mesh refinement would require about 20 levels of refinement to cover a similar range of scales, assuming refinements by factors of two between each level.  

We impose ``moving puncture" coordinate condition during our evolution calculations.  Specifically, we adopt the 1+log slicing condition \cite{BonMSS95}
\begin{equation} \label{onepluslog}
(\partial_t - \beta^i \partial_i) \alpha = - 2 \alpha K
\end{equation}
for the lapse, where $K$ is the mean curvature, and choose a Gamma-driver condition \cite{AlcBDKPST03} for the shift, namely the version presented in \cite{ThiBB11}. 

\subsection{Diagnostics}
\label{sec:diagnose}

We use several diagnostics to analyze the properties of our numerical solutions.  

As discussed in Section \ref{sec:numerics}, we allowed regridding to small values of the outer boundary in order to be able to follow the self-similar solution to late times.  The disadvantage of this approach is that we had to terminate the simulations before the horizons of newly formed black holes had enough time to settle down to approximate equilibrium.  In this paper we therefore do not analyze the scaling of black hole masses for supercritical data.  As pointed out by \cite{GarD98}, however, scaling can also be observed for subcritical data.  The maximum encountered value of the central density $\rho_c$, for example, has units of inverse length, and therefore must satisfy a scaling law
\begin{equation} \label{rho_scaling_overall}
\rho_c^{\rm max} \simeq |\eta_* - \eta|^{-2 \gamma}.
\end{equation}
We note that the density $\rho$, defined in (\ref{rho}), is in general not an invariant quantity, since it depends on the spacetime slicing through the normal vector $n^a$.  At the origin of the coordinate system, however, the symmetries in the problem single out a preferred normal vector, so that the values of the central density $\rho_c$ reported here do take on an invariant meaning.

The scaling laws (\ref{mass_scaling}) and (\ref{rho_scaling_overall}) describe the overall behavior of quantities close to criticality, but both Gundlach \cite{Gun97} and Hod and Piran \cite{HodP97b} realized that, for scalar fields, the discretely self-similar nature of the critical solution imprints a periodic fine structure on the scaling law.  The scaling for the maximum encountered central density can then be written as
\begin{equation} \label{rho_scaling_0}
\ln \rho_c^{\rm max} = C - 2 \gamma \ln|\eta_* - \eta| + f(\ln|\eta_* - \eta|),
\end{equation}
where $C$ is the logarithm of the constant of proportionality between $\rho_c$ and the term $|\eta_* - \eta|^{-2\gamma}$ in (\ref{rho_scaling_overall}), and where $f(x)$ is a function that is periodic in $x$ with angular frequency
\begin{equation} \label{omega}
\omega = \frac{\Delta}{2 \gamma}.  
\end{equation}
In the fits presented in Section \ref{sec:results} we assume that, to leading order, we can approximate $f(x)$ as
\begin{equation}
f(x) = A \sin(\omega x + \phi_{\rm ph}),
\end{equation}
where $A$ is the amplitude of the fine structure and $\phi_{\rm ph}$ a phase, so that (\ref{rho_scaling_0}) becomes
\begin{equation} \label{rho_scaling}
\ln \rho_c^{\rm max} = C - 2 \gamma \ln|\eta_* - \eta| + A \sin(\omega \ln|\eta_* - \eta| + \phi_{\rm ph}).
\end{equation}
Fits to numerical data then provide values for $\eta_*$, the critical exponent $\gamma$, as well as $\omega$.  Using (\ref{omega}) we can then compute $\Delta$ from $\omega$ and $\gamma$ -- the first of three different approaches to computing the period $\Delta$.

We also adopt an approach similar to that of CHLP to monitor the departure from sphericity.  At logarithmic times $T_0$ (see Eq.~\ref{T}) we launch pairs of photons from the center, one along the axis ($\theta = 0$) and one in the equatorial plane ($\theta = \pi/2$).   For each pair of photons we initialize the affine parameter $\lambda$ to zero, and the derivative of $\lambda$ to that of $T$, i.e. $d \lambda/dt  = dT/dt$ (other normalizations are possible, but with this normalization $\lambda$ inherits the self-similar nature of $T$).  We follow the photons' trajectories during the subsequent evolution, and record the value of the scalar field $\phi$ at each photon's current location as a function of its affine parameter $\lambda$.  We then compute the difference
\begin{equation} \label{delta_phi}
\delta \phi(T_0, \lambda) = 
\phi_{\rm ax}(\lambda) - \phi_{\rm eq}(\lambda)
\end{equation}
for a pair of photons launched at $T_0$, as well as the maxima of these differences
\begin{equation} \label{Delta_phi}
\Delta \phi (T_0) \equiv \max_{\lambda} | \delta \phi(T_0, \lambda) |,
\end{equation}
for each pair of photons emitted at logarithmic time $T_0$.  

For $\epsilon^2 > 0$, a projection of our initial data (\ref{indata}) into Legendre polynomials would result in contributions to modes of all even order $\ell$.  However, we expect higher-order modes to decay more rapidly than lower-order modes (see Table~I in MGG) so that at sufficiently late times our diagnostics $\delta \phi$ and $\Delta \phi$ become a measure of $\ell = 2$ modes.

\section{Results}
\label{sec:results}

\subsection{Spherical symmetry}
\label{sec:spherical}

In order to both test and calibrate our code, and for better comparison with our aspherical results in Section \ref{sec:deformations}, we start with a discussion of results in spherical symmetry, i.e.~for $\epsilon = 0$ in the initial data (\ref{indata}).  Essentially the results presented in this Section reproduce those of \cite{Cho93} and numerous follow-up publications, including \cite{AkbC15} who, similar to our treatment here, adopted the BSSN formalism and moving-puncture coordinates. 

\begin{figure}[t]
\begin{center}
\includegraphics[width=3in]{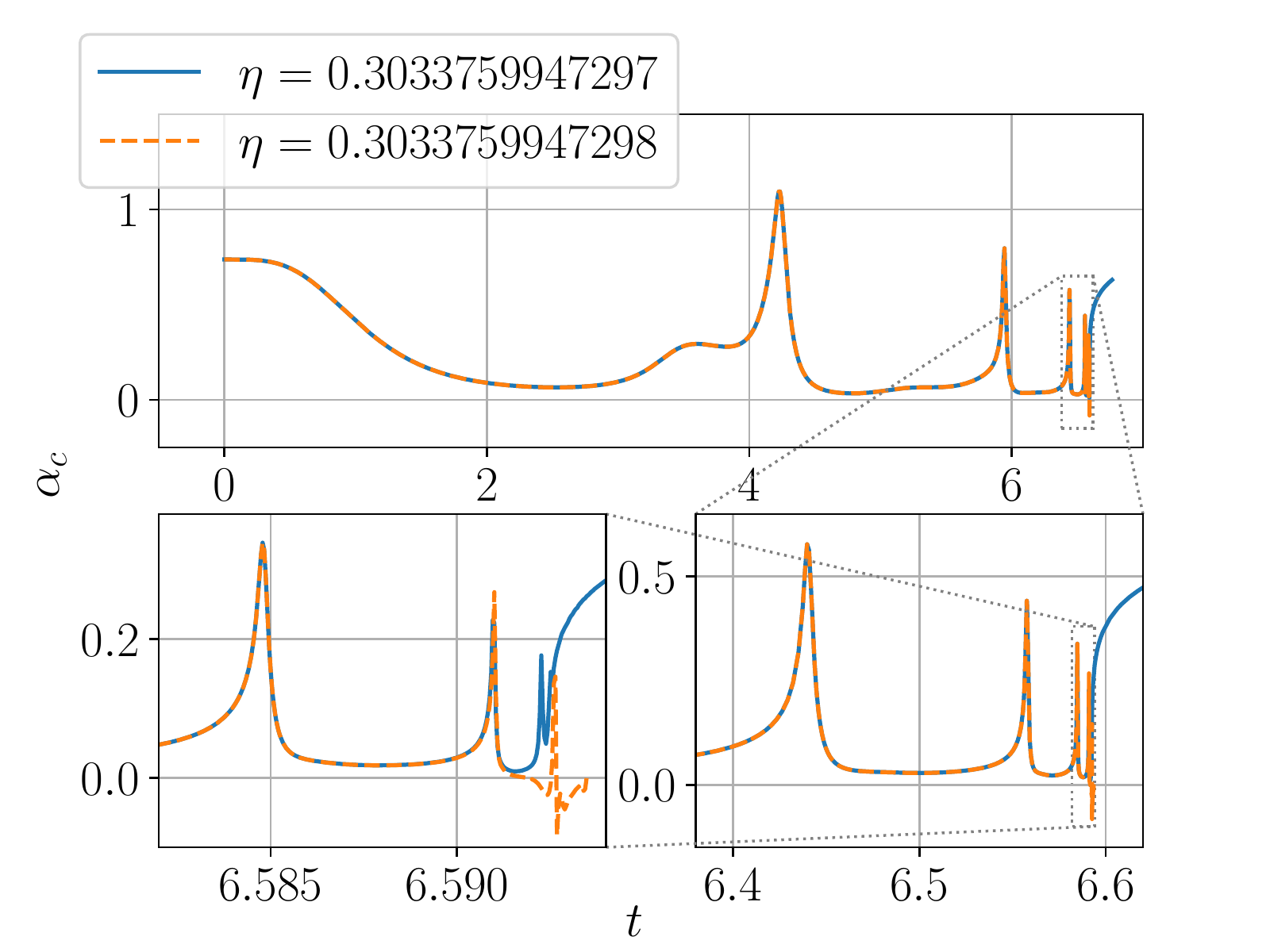}
\end{center}
\caption{The central value of the lapse function $\alpha$ as a function of coordinate time $t$ for two spherically symmetric evolutions with values of $\eta$ that bracket the critical value $\eta_*$.}
\label{fig:lapse_spherical}
\end{figure}

We bracket the critical parameter $\eta_*$ using bisection, except that in each step we refine the bracketing interval into ten new intervals instead of two by adding one more digit to $\eta$.   For subcritical data the scalar field disperses, leaving behind flat space, and the lapse function $\alpha$ approaches unity, while for supercritical data the minimum of the lapse function drops to zero, indicating the formation of a black hole.  We show an example in Fig.~\ref{fig:lapse_spherical}, which already suggests the presence of a pattern that keeps repeating on increasingly small scales.

While we can bracket the critical parameter $\eta_*$ to about 13 significant digits, we would like to emphasize that these numbers do depend on the numerical grid.  The true uncertainty in $\eta_*$ is significantly larger than the 13 digits suggest; see our discussion below.

\begin{figure}[t]
\begin{center}
\includegraphics[width=3in]{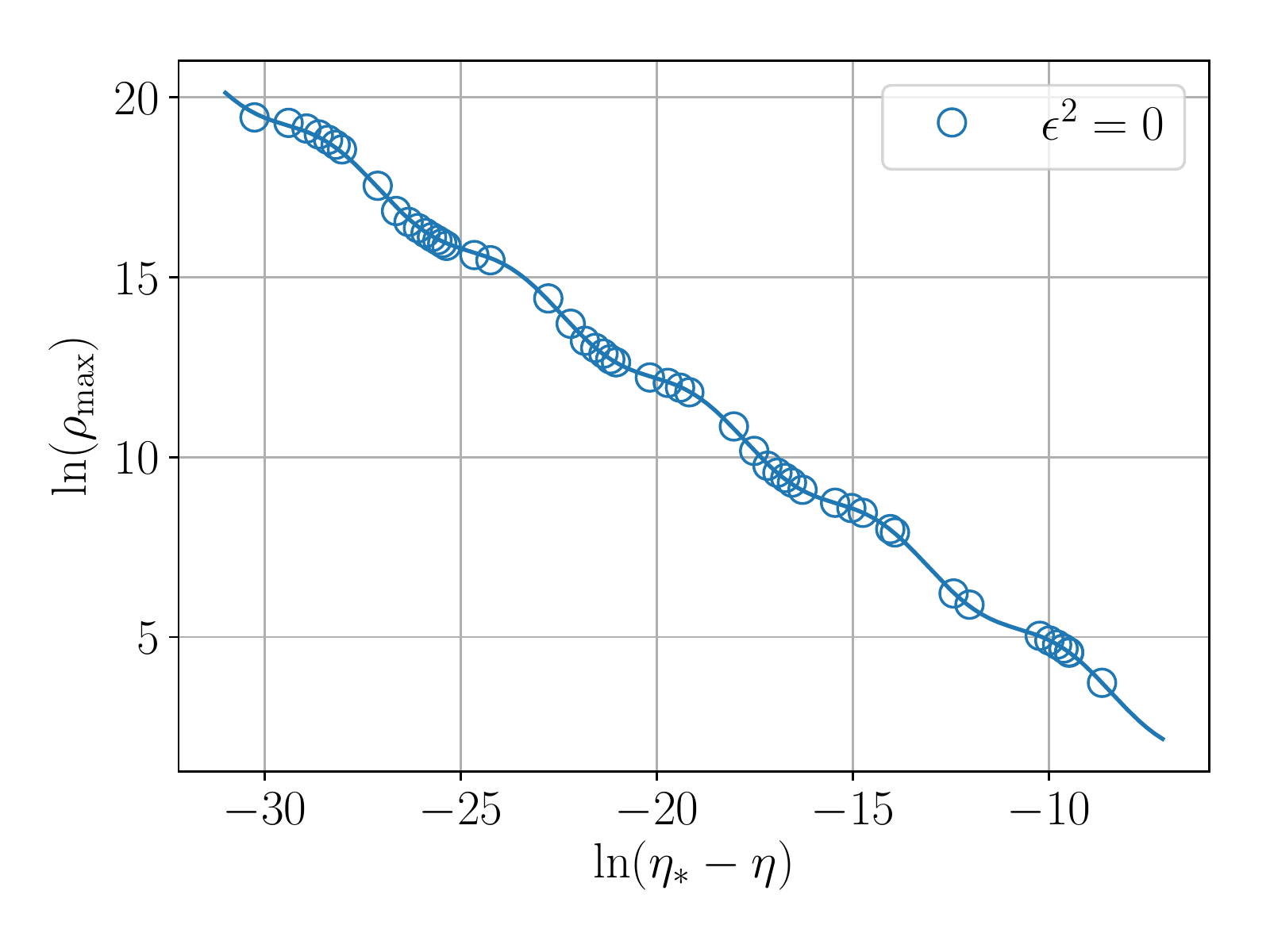}
\end{center}
\caption{Scaling of the maximum central density $\rho_c^{\rm max}$ for spherically symmetric subcritical data.  Open circles represent numerical results from spherical evolutions, while the solid line is a fit based on (\ref{rho_scaling}).}
\label{fig:rho_max_spherical}
\end{figure}

In Fig.~\ref{fig:rho_max_spherical} we show numerical results for the maximum encountered central density $\rho_c$ as a function of $\eta$ for subcritical data.   We fit the numerical data to (\ref{rho_scaling}) and obtain values for $\eta_*$, $\gamma$ and $\omega$ (as well as $C$, $A$ and $\phi_{\rm ph}$).  From (\ref{omega}) we can then obtain our first estimate of the echoing period $\Delta$. 

Several different sources of error affect the uncertainty in the parameters determined in these fits.   The critical amplitude $\eta_*$ is mostly affected by numerical finite-difference error.  Based on comparisons with different grid setups we  expect this value to be accurate to within about 0.1\% for the spherically symmetric simulations considered here.  The coefficients $\gamma$ and $\Delta$ also depend more sensitively on which data are included in the fits -- evolutions too close to the critical point develop features that can no longer be resolved on our grids, and therefore lead to large errors, while for evolutions far way from the critical point the expected power-law scalings no longer apply.  We crudely estimate the resulting errors to be about 1\%; slightly less for $\gamma$, and slightly more for $\Delta$.

\begin{figure}[t]
\begin{center}
\includegraphics[width=3in]{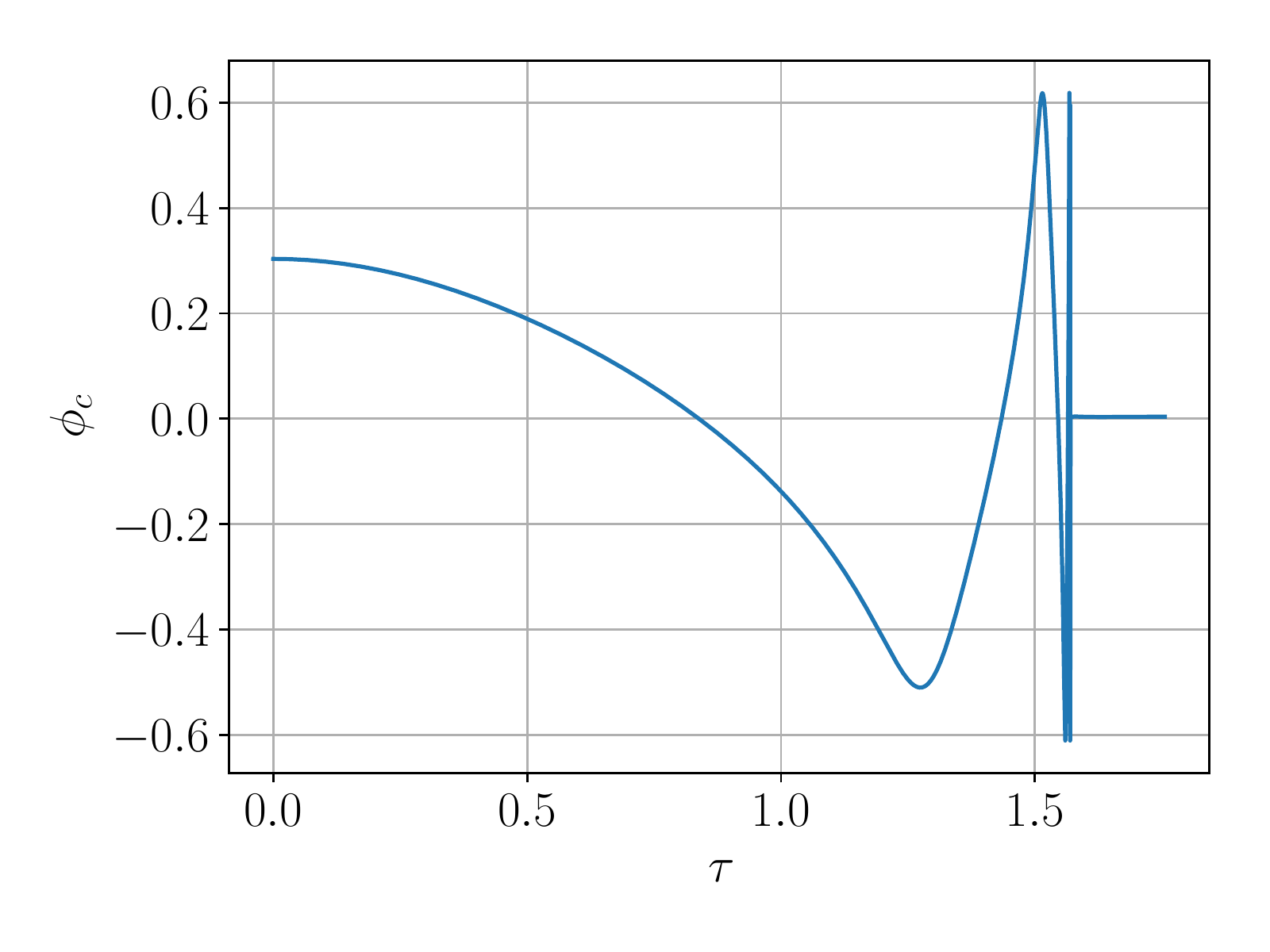}
\end{center}
\caption{The central value of the scalar field $\phi$ versus proper time $\tau$ for a near-critical, spherically symmetric evolution.  The field displays oscillations with decreasing period, as expected for a self-similarly contracting solution, accumulating at the accumulation time $\tau_*$.} 
\label{fig:phi_central_spherical_tau}
\end{figure}

\begin{figure}[t]
\begin{center}
\includegraphics[width=3in]{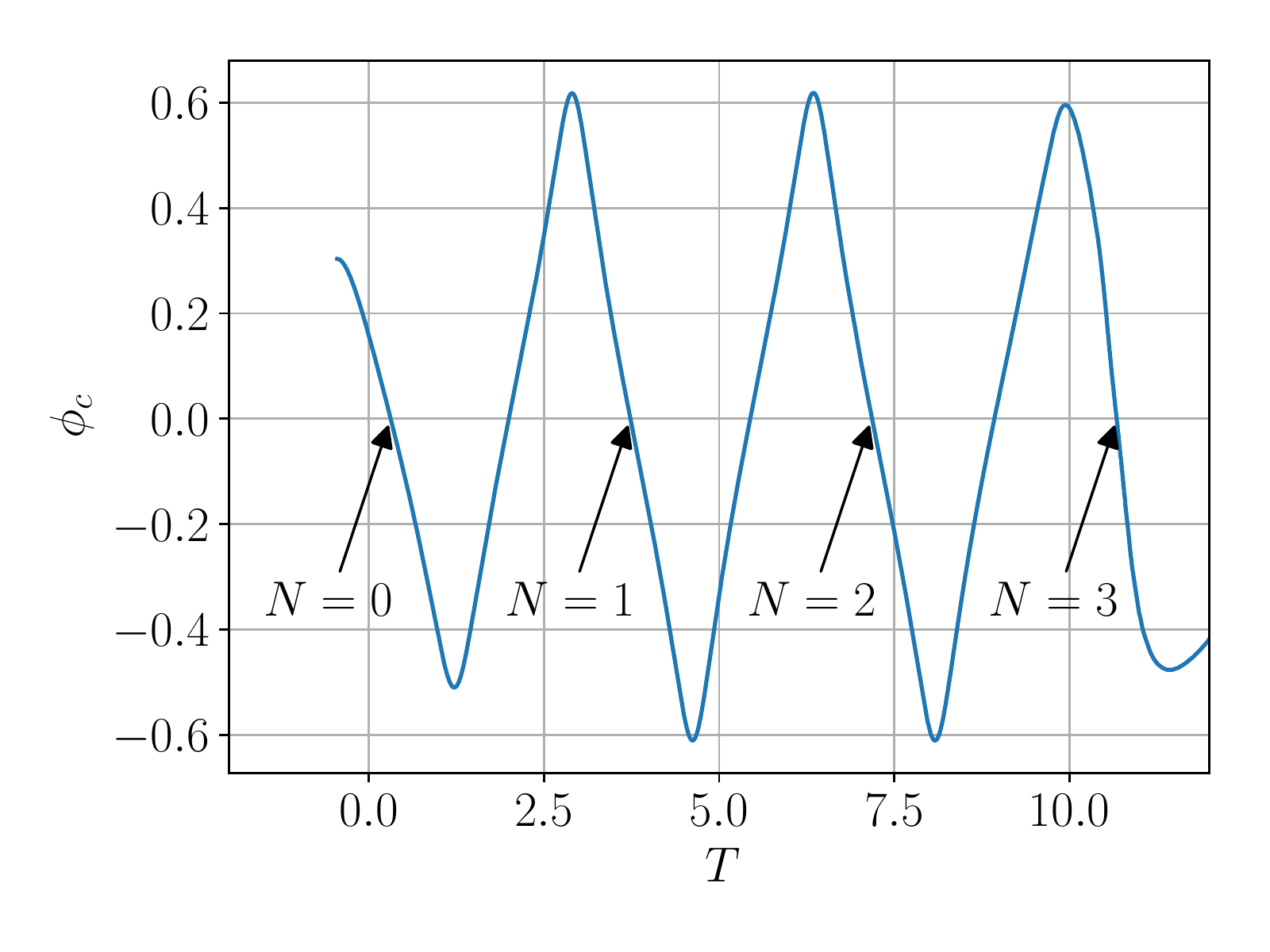}
\end{center}
\caption{Same as Fig.~\ref{fig:phi_central_spherical_tau}, but plotted as a function of $T$ rather than $\tau$.  The labels identify every other zero-crossing of $\phi_c$; in Fig.~\ref{fig:echoes_spherical} we show radial profiles of $\phi$ at these times in order to demonstrate echoing.  (See also Table \ref{table:zeros} for the coordinate and proper times of these zero-crossings.)}
\label{fig:phi_central_spherical}
\end{figure}

We next consider the central value $\phi_c$ of the scalar field for a subcritical evolution close to the critical point.  When plotted as a function of proper time of an observer at the origin, as in Fig.~\ref{fig:phi_central_spherical_tau}, $\phi_c$ displays oscillations with decreasing period, as one would expect for a self-similarly contracting solution whose time and length scales continuously decrease.  The oscillations accumulate at the accumulation time $\tau_*$.  

We can determine the accumulation time $\tau_*$ by looking for periodic behavior in the logarithmic time $T$ (see Eq.~(\ref{T})).  In practice we identify the proper times $\tau_n$ of zero-crossings of $\phi_c$.  We then consider a pair of subsequent zero-crossings, say $\tau_{n}$ and $\tau_{n+1}$, and a second pair $\tau_m$ and $\tau_{m+1}$.  Assuming that the advance of logarithmic time $T$ for each pair is equal -- namely $\Delta/2$ -- results in the estimate
\begin{equation} \label{tau_star}
\tau_* = \frac{\tau_n \tau_{m+1} - \tau_{n+1} \tau_m}{\tau_n - \tau_{n+1} - \tau_m + \tau_{m+1}}.
\end{equation}
Equating the advance of $T$ for each pair to $\Delta/2$ then yields 
\begin{equation} \label{Delta_period}
\Delta = 2 \ln \left( \frac{\tau_* - \tau_n}{\tau_* - \tau_{n+1}} \right),
\end{equation}
our second approach to determining $\Delta$.  In practice we compute $\Delta$ for all pairs during the self-similar part of the evolution, and report average values in Table \ref{table:results}.  For different pairs, the values of $\Delta$ differ by up to approximately 1\%, which we therefore take to be the approximate error in this value.   In Fig.~\ref{fig:phi_central_spherical} we show the central value of the scalar field as a function of $T$, which clearly reveals the periodic behavior during an intermediate regime.  

\begin{table}
\begin{tabular}{l|c|c|c|c|c|c}
\hline 
\hline
 			& $\eta_*$ 	& $\tau_*$	 	& $\gamma$  	&  \multicolumn{3}{c}{$\Delta$} \\
$\epsilon^2$ 	&			&			&			& (\ref{omega}) & (\ref{Delta_period}) & (\ref{R_echoes}) \\
\hline \hline
quasi-analytical	& --			& --			& 0.374		& \multicolumn{3}{c}{3.445452402} \\
\hline
0			& 0.303376	& 1.570		& 0.374		&  3.47		& 3.46  	& 3.46 \\
0.01			& 0.303352	& 1.572		& 0.374		&  3.47		& 3.47 	& 3.45 \\ 	
0.5			& 0.304512	& 1.775		& 0.369		&  3.39		& 3.37  	& 3.35 \\
0.75			& 0.308378	& 2.067		& 0.306		&  2.87		& 2.80 	& 2.75 	\\
\hline
\hline
\end{tabular}
\caption{Summary of our numerical results.  The ``quasi-analytical" results (from \cite{Gun95,Gun97,MarG03}) provide values for $\gamma$ and $\Delta$, but not for $\eta_*$ or $\tau_*$, which depend on the specifics of the initial data.  We report values of $\Delta$ as computed from three different approaches, namely from the frequency of the fine-structure of the critical scaling, i.e.~Eq.~(\ref{omega}), from the periodicity of the scalar field at the origin, Eq.~(\ref{Delta_period}), and from the scaling of radial echoes, Eq.~(\ref{R_echoes}).   See text for estimates of our errors;  in particular, $\Delta$ can be determined only very crudely from the scaling of the echoes for large values of $\epsilon$.  We find good agreement with the perturbative results for small departures from sphericity, but find that both $\gamma$ and $\Delta$ decrease as $\epsilon^2$ increases (compare Table I in CHLP).}  
\label{table:results}
\end{table}

Fig.~\ref{fig:phi_central_spherical} helps distinguish three different phases during the evolution.  In Phase 1, the initial data approach the self-similar, critical solution.  This phase depends on the specifics of the initial data, but with sufficient fine-tuning it will result in the critical solution, the Choptuik spacetime, plus a perturbation whose amplitude depends on the degree of fine-tuning.     During Phase 2, the evolution can then be described as the critical solution plus the slowly growing perturbation.  Once the perturbation has become sufficiently large, the evolution departs from critical solution nonlinearly, marking the transition from Phase 2 to Phase 3, and the scalar field either disperses to infinity or collapses into a black hole.  The intermediate regime during which the periodic behavior can be observed in Fig.~\ref{fig:phi_central_spherical} corresponds to Phase 2.

\begin{figure}[t]
\begin{center}
\includegraphics[width=3in]{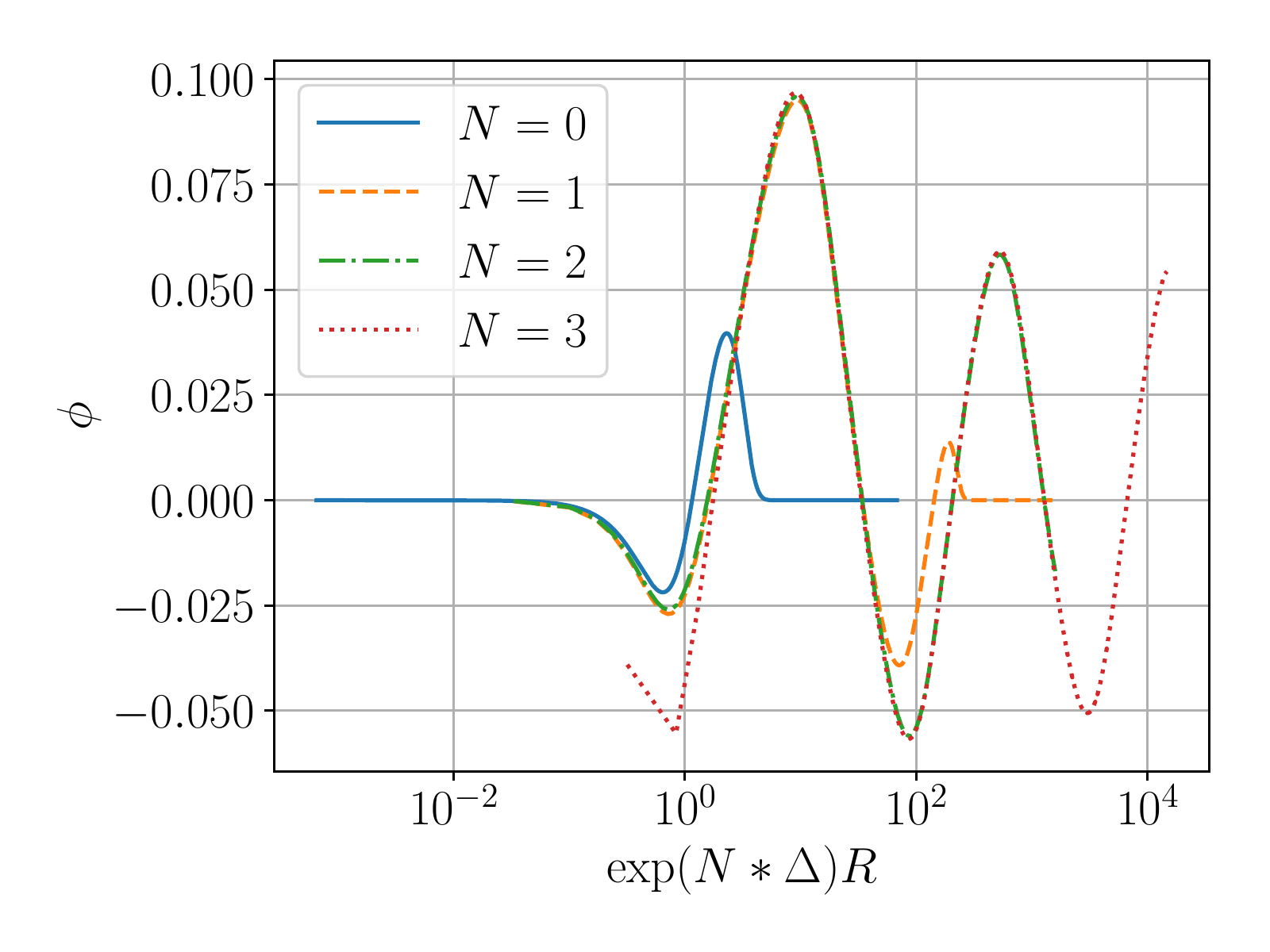}
\end{center}
\caption{Radial profiles of $\phi$ at the times identified in Fig.~\ref{fig:phi_central_spherical}, i.e.~at every other zero-crossing of the central value $\phi_c$ (see also Table \ref{table:zeros} for coordinate and proper times of these zero-crossings).  We show $\phi$ as a function of proper distance $R$ from the origin, on a logarithmic scale, and the $N$-th echo rescaled with a factor $e^{N\Delta}$ with $\Delta = 3.45$ (see Eq.~\ref{R_echoes}; note that $e^{3 \Delta} \simeq 31,000$).  In regimes for which the solution has entered the self-similar phase, and for which we have not yet lost numerical resolution close to the origin, the radial profiles agree very well, clearly demonstrating echoing.}
\label{fig:echoes_spherical}
\end{figure}

In Fig.~\ref{fig:phi_central_spherical} we also label every other zero-crossing of $\phi_c$, i.e.~times $T$ that, during Phase 2, are separated by a whole period $\Delta$, as opposed to the half-periods discussed above.  We list both coordinate times and proper times of these zero-crossings in Table \ref{table:zeros}.    Note that $N = 0$ occurs just before the evolution enters Phase 2, while for $N=3$ the resolution at the center becomes quite poor.  According to the properties of the discretely self-similar solution, radial profiles of $\phi$ at these times should be echoes of each other, meaning that they should be identical to each other (in those parts of the solution that have reached self-similarity already), except rescaled by a factor $e^{-\Delta}$ between each echo (see Eq.~\ref{R_echoes}).  In Fig.~\ref{fig:echoes_spherical} we show these profiles as a function of proper distance $R$ from the origin, which we find by integrating from the origin to each grid point along lines of constant angles $\theta$ and $\varphi$.  For those parts of the profiles in the self-similar regime (and with sufficient numerical resolution) we find excellent agreement if we choose $\Delta = 3.45$.  This provides us with a third measure of the echoing period $\Delta$.   The agreement becomes visually worse if we increase or decrease $\Delta$ by about 1\%.  

As an aside we note that it was not clear a priori that the discretely self-similar symmetry of the Choptuik spacetime would be revealed in our time slicing, but apparently the 1+log slicing (\ref{onepluslog}) does reflect the symmetry of the spacetime (see also the discussion in \cite{CelB18} in the context of ultra-relativistic fluids; see also \cite{GarG99} for proposals for symmetry-seeking spacetime coordinates).  

\begin{table}
\centering
\begin{tabular}{l|l|l|l|l|l|l}
\hline
\hline
		 & \multicolumn{2}{c|}{$\epsilon^2 = 0$} & \multicolumn{2}{c|}{$\epsilon^2 = 0.5$} & \multicolumn{2}{c}{$\epsilon^2 = 0.75$} \\ 
\hline
$N$		   & $t$ 	& $\tau$ & $t$ 	& $\tau$ & $t$ 	& $\tau$ \\
\hline
\hline
 0	& 1.85344	 	& 0.838213  	&  2.15922 	&	0.943157 	& 2.70846	& 1.06692 \\
 1	& 6.27193  	& 1.54598 	&  7.35928	&	1.74628 	& 9.54811	& 2.01083 \\
 2	& 6.57627		& 1.56913 	&  7.7259		&	1.77395 	& 10.1933	& 2.06492 \\
 3	& 6.59199		& 1.56987 	&  7.7457		&	1.77491	&  10.2777 & 2.07448 \\
\hline
\hline
\end{tabular}
\caption{Coordinate times $t$ and proper times $\tau$ (as measured by an observer at the origin) of the $N$-th other zero-crossing of the scalar field $\phi$ at the origin (as indicated in Fig.~\ref{fig:phi_central_spherical}).  The number of digits provided does not reflect the numerical error in these data. }
\label{table:zeros} 
\end{table}

We report all our results in Table \ref{table:results}.  In particular we find that our different approaches to determining $\Delta$ yield values that are within our estimated margins of error.  Our values are also in good agreement with most previous numerical studies (e.g.~\cite{AkbC15}) as well as the values provided by \cite{Gun95,Gun97,MarG03}, who found $\Delta$ by casting the problem as an eigenvalue problem.  Similarly, our values of the critical exponent agree well with previous numerical results, as well as the perturbative values found by \cite{Gun97}.  In the following, including in Table \ref{table:results}, we will refer to the results of \cite{Gun95,Gun97,MarG03} as ``quasi-analytical" results, even though solving the eigenvalue problem or the perturbative equations involves numerical work.

\subsection{Aspherical deformations}
\label{sec:deformations}

\subsubsection{Critical parameters}

We next consider aspherical evolutions with $\epsilon > 0$ in the initial data (\ref{indata}).   For a given value of $\epsilon$ we bracket the critical value $\eta_*$ as described in Section \ref{sec:spherical}.  

\begin{figure}[t]
\begin{center}
\includegraphics[width=3in]{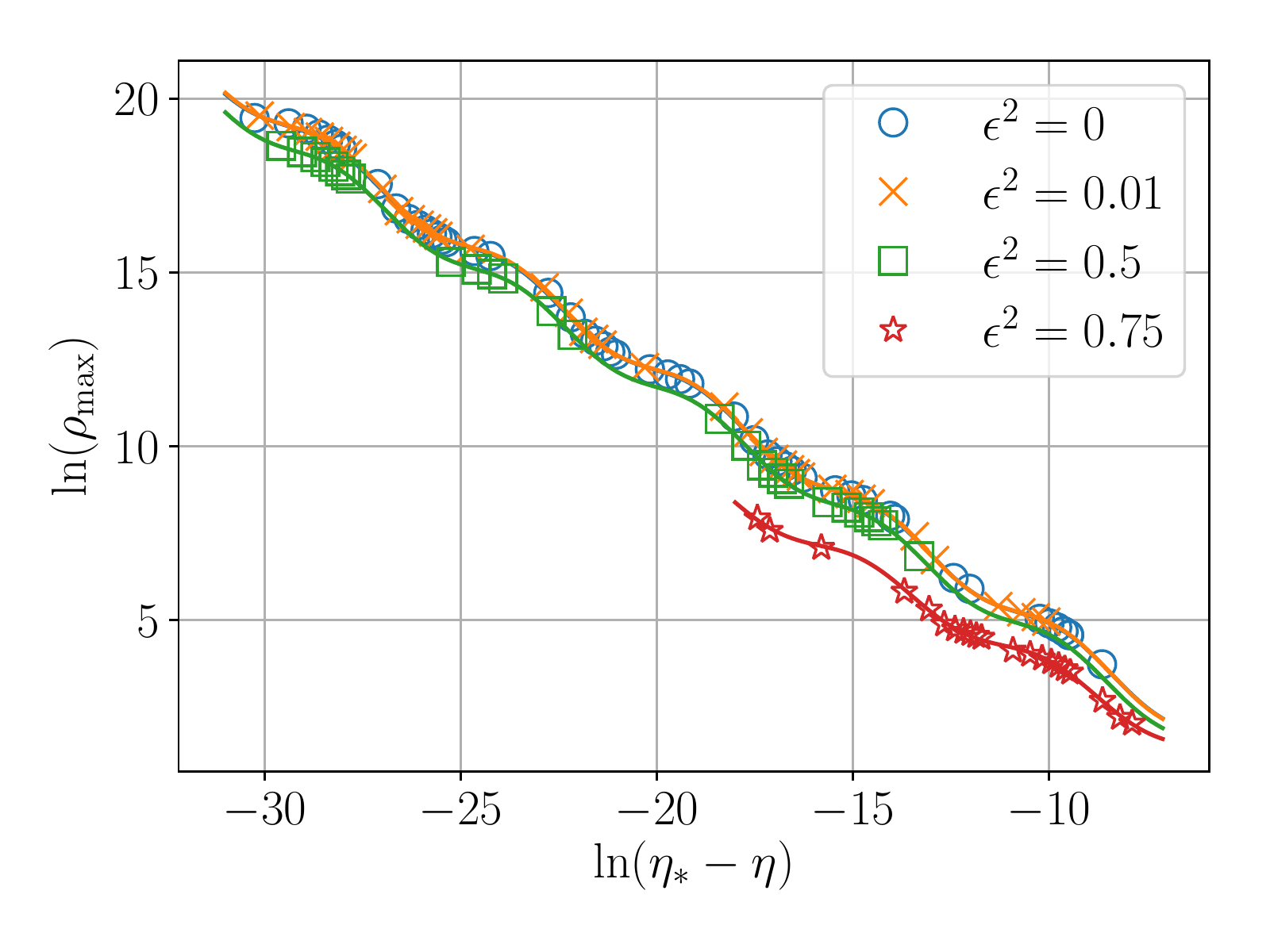}
\end{center}
\caption{Same as Fig.~\ref{fig:rho_max_spherical}, but now including results for aspherical evolutions with $\epsilon > 0$.  The symbols represent numerical results, and the solid lines fits based on (\ref{rho_scaling}).}
\label{fig:rho_max}
\end{figure}

\begin{figure}[t]
\begin{center}
\includegraphics[width=3in]{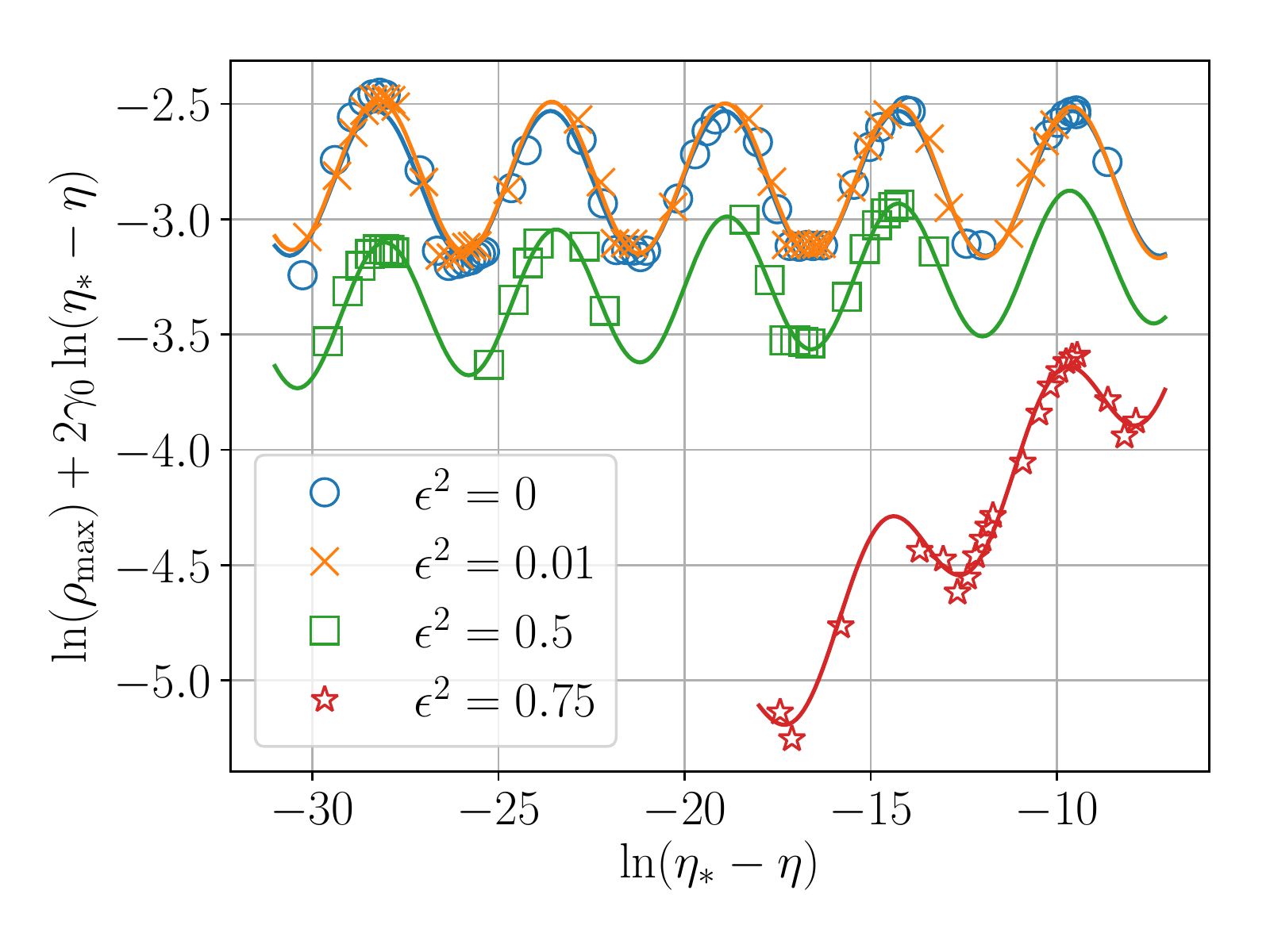}
\end{center}
\caption{Same as Fig.~\ref{fig:rho_max}, but with the overall scaling $- 2 \gamma_0 \ln(\eta_* - \eta)$ subtracted in order to highlight the difference between the spherical and aspherical scalings (see Eq.~\ref{rho_scaling}).  Here $\gamma_0 = 0.374$ is the value of the critical exponent found in our spherical evolutions.  (Compare Fig.~3 in CHLP.)}
\label{fig:rho_max_rescaled}
\end{figure}

For subcritical evolutions we can again fit the values of the maximum encountered density to the scaling relation (\ref{rho_scaling}).  The result of these fits is shown in Fig.~\ref{fig:rho_max}, which generalizes Fig.~\ref{fig:rho_max_spherical} to include aspherical evolutions.  The overall trend - namely that the critical exponent $\gamma$ appears to decrease with increasing $\epsilon$ -- can already be observed in Fig.~\ref{fig:rho_max}, but it is easier to see this after subtracting the overall scaling $- 2 \gamma_0 \ln(\eta_* - \eta)$  (see Eq.~\ref{rho_scaling}) from each curve.   Here $\gamma_0 = 0.374$ is the value that we found in our spherical evolutions (see Table \ref{table:results}).  The resulting curves are shown in Fig.~\ref{fig:rho_max_rescaled} -- note the similarity with the corresponding Fig.~3 in CHLP.   For $\epsilon^2 = 0.75$ we included data only for relatively large values of $\ln(\eta_* - \eta)$.   Closer to the black-hole threshold we found larger deviations from these fits, which we believe are caused by a growing aspherical mode, as we will discuss in more detail below.

From the fits to (\ref{rho_scaling}) we determine both $\gamma$ and, via (\ref{omega}), the echoing period $\Delta$ (see Table \ref{table:results}).   For small deformations, we estimate the errors to be similar to those that we found for spherically symmetric data.  For larger values of $\epsilon$, two effects lead to increasing errors.  A stronger angular dependence leads to larger numerical finite-difference errors for a given angular resolution.  Moreover, for $\epsilon^2 = 0.75$ we included fewer data points in the fits, as discussed above, which also increases the error.  We estimate the latter by varying the number of points included in the fits and find that, for large $\epsilon$, the errors in $\gamma$ and $\Delta$ are closer to about 5\%.  We nevertheless note that both values decrease with increasing $\epsilon$, in good agreement with the findings of CHLP (see their Table~I).   

The critical exponent $\gamma$ and the echoing period $\Delta$ describe properties of the spherically symmetric critical solution and its linear perturbations; strictly speaking, therefore, they take unique values that are independent of $\epsilon$.   Apparently, however, the dynamics of nonlinear perturbations of the critical solution can still be described heuristically in terms of similar parameters $\gamma$ and $\Delta$, except that they now take on effective values that do depend on $\epsilon$.  When we discuss the dependence of $\gamma$ and $\Delta$ on $\epsilon$ in the following, we mean these effective values, rather than those defined in the context of linear perturbation theory.

\begin{figure}[t]
\begin{center}
\includegraphics[width=3in]{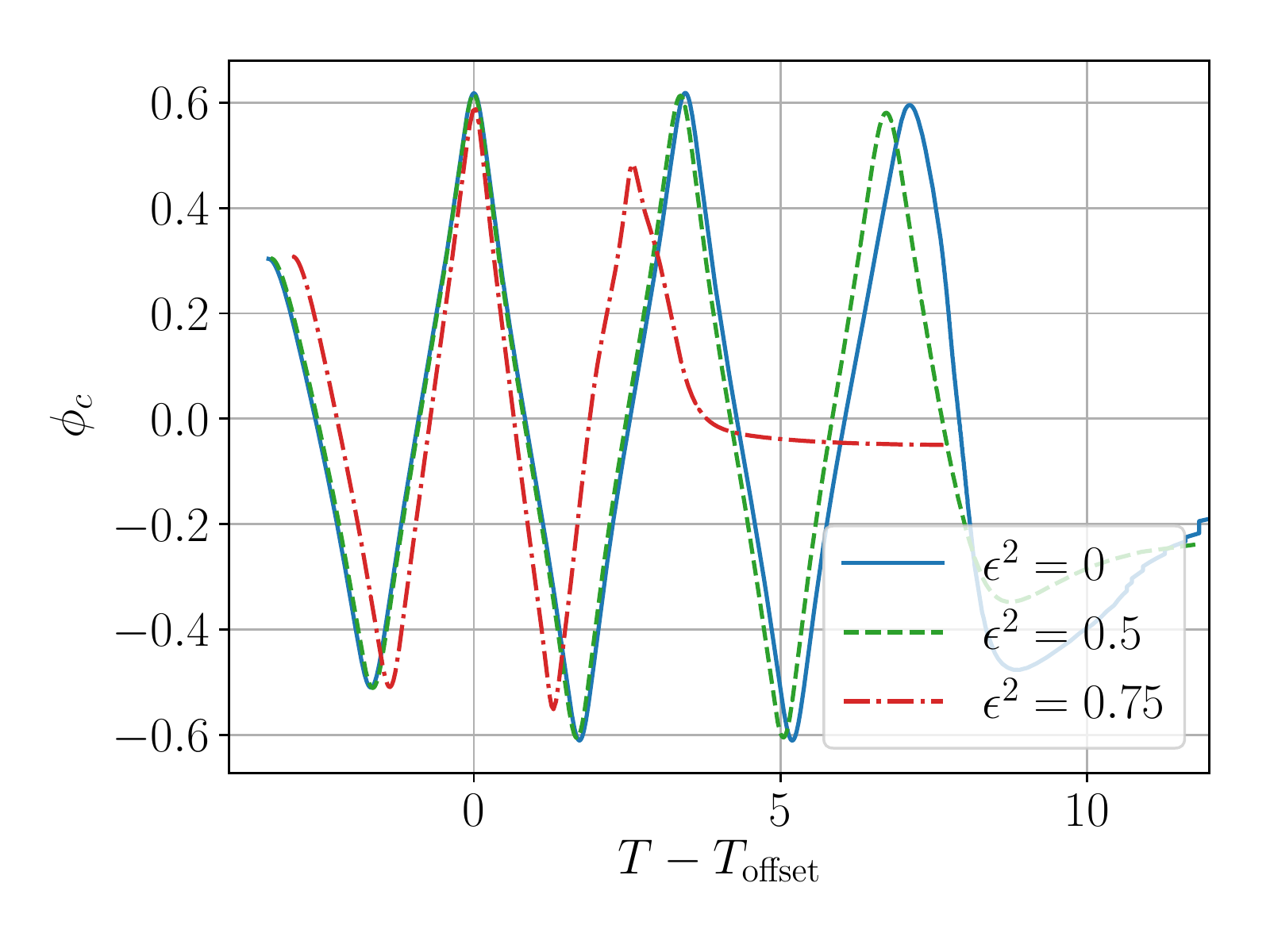}
\end{center}
\caption{Same as Fig.~\ref{fig:phi_central_spherical}, but now including aspherical near-critical evolutions for different values of $\epsilon$.  We also introduced an offset $T_{\rm offset}$, defined as the logarithmic time $T$ of the first maximum, for an easier comparison of the echoing period.  (Compare Fig.~4 in CHLP.)}  
\label{fig:phi_central} 
\end{figure}

As for our spherically symmetric evolutions in Section \ref{sec:spherical}, we next consider the time dependence of the central value of $\phi$ for a near-critical evolution.   Identifying the proper times of zero-crossings during Phase 2 we can determine the accumulation time $\tau_*$ from (\ref{tau_star}), and the echoing period $\Delta$ from (\ref{Delta_period}).  Our results for different values of $\epsilon$ are listed in Table~\ref{table:results}.  For $\epsilon^2 = 0.01$ and $0.5$ we estimate the error in $\Delta$ to be again approximately 1\%.  For $\epsilon^2 = 0.75$, however, we observe fewer zero-crossings during Phase 2, and therefore believe that our error is close to about 5\%.   As before we find that $\Delta$ decreases with increasing $\epsilon$, in good agreement with our previous determination of $\Delta$ from the scaling of $\rho_c^{\rm max}$.  In Fig.~\ref{fig:phi_central} we show $\phi_c$ as a function of logarithmic time $T$ for different values of $\epsilon$; this graph also shows that the period $\Delta$ becomes smaller for increasing $\epsilon$.  

\begin{figure}[t]
\begin{center}
\includegraphics[width=3in]{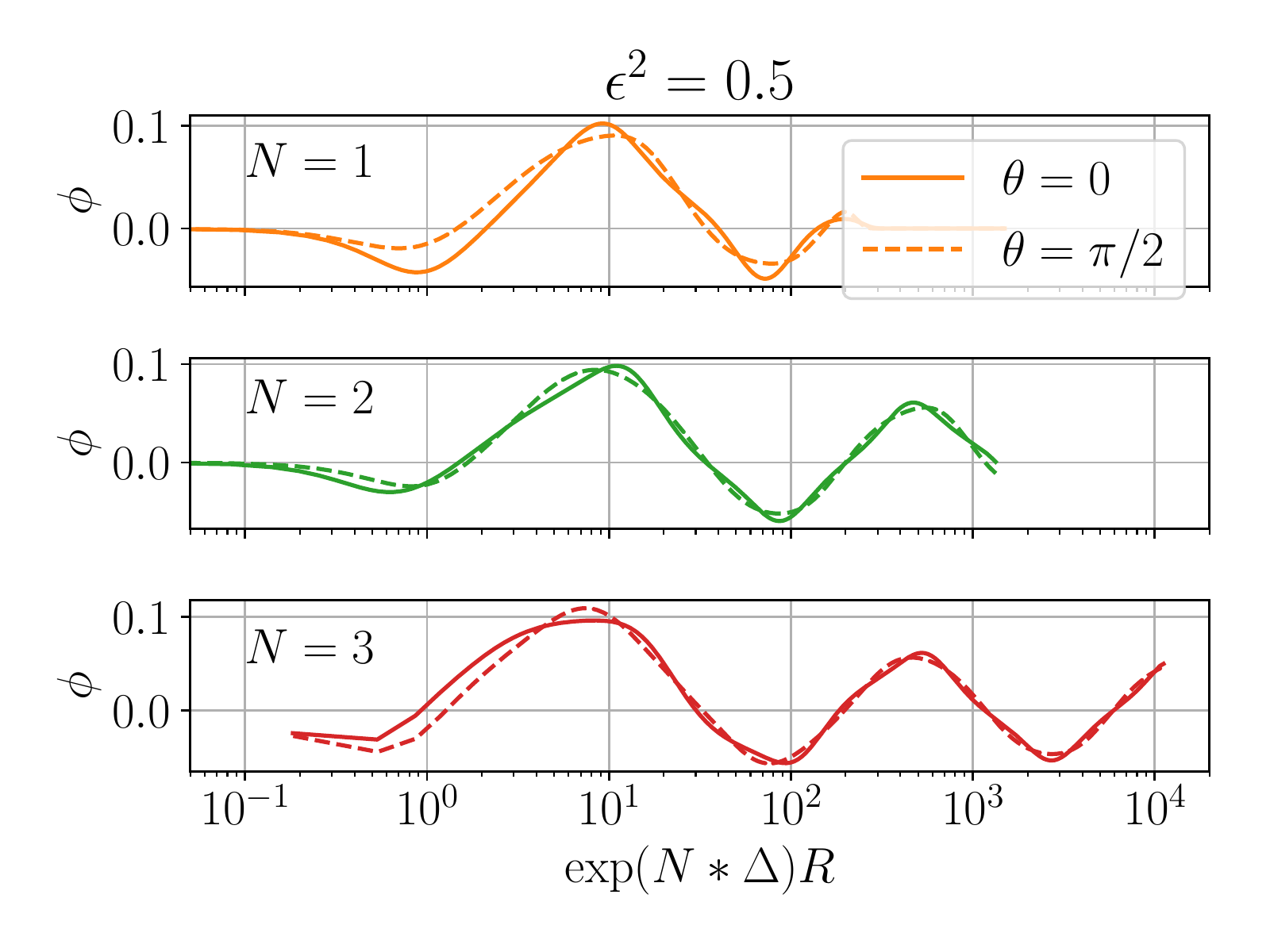}
\end{center}
\caption{Echoing in spatial profiles of the scalar field $\phi$ for $\epsilon^2 = 0.5$, with $\Delta = 3.35$.}
\label{fig:echoes_05}
\end{figure}

\begin{figure}[t]
\begin{center}
\includegraphics[width=3in]{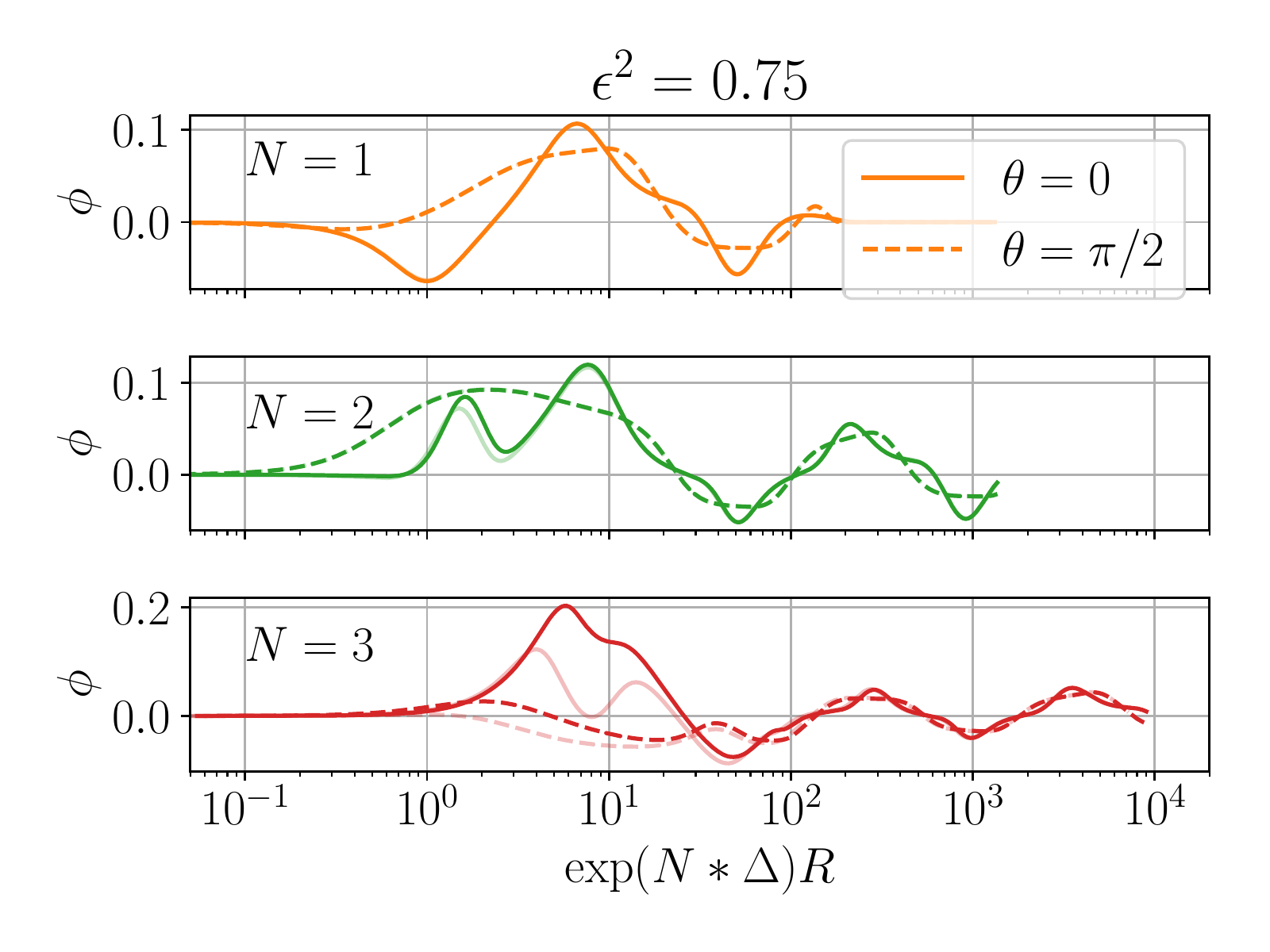}
\end{center}
\caption{Same as Fig.~\ref{fig:echoes_05}, but for $\epsilon^2 = 0.75$ and with $\Delta = 2.75$.  At the time of the third echo a new feature has emerged on the axis (note the different scale on the $y$-axis for $N = 3$).  We include results for $N_\theta = 14$ (solid colors) as well as $N_\theta = 12$ (faded) colors.  At early times, as well as at sufficiently large radii, the two cannot be distinguished in this plot.  While the increasing angular dependence of the emerging feature leads to increasing quantitative differences, we nevertheless find very similar qualitative behavior independently of angular resolution (see also Fig.~\ref{fig:bifurcation} as well as the discussion in the text).}
\label{fig:echoes_075}
\end{figure}

Next we again consider echoing in radial profiles of the scalar field $\phi$ at moments of alternating zero-crossings of its central value $\phi_c$.   We again list the times of these zero-crossings in Table \ref{table:zeros}.  The echoing is now harder to identify than for the spherical evolutions, because the new angular-dependency leads to additional oscillations with periods different from $\Delta$, meaning that the echoing is no longer exact (see also Eq.~\ref{perturb} below).  For small values of $\epsilon$ the overall time dependence of $\phi$ is still dominated by that of the spherical Choptuik spacetime, so that the echoing can still be seen very cleanly, and $\Delta$ can still be determined quite accurately.  For $\epsilon^2 = 0.5$ the deviations are larger, as we show in Fig.~\ref{fig:echoes_05}.  We now show each echo in its own panel, and include profiles of $\phi$ both in the axial direction ($\theta = 0$, solid lines) and the equatorial direction ($\theta = \pi/2$, dotted lines) in each panel.   While the profiles in the two directions do not agree, the differences do not appear to decay or grow with time (i.e.~from echo to echo).  Moreover we can still identify the overall features of the Choptuik spacetime, which match quite well if we rescale each echo with $\Delta = 3.35$.

The behavior becomes more complicated for $\epsilon^2 = 0.75$, as shown in Fig.~\ref{fig:echoes_075}.  It now appears that the differences between the axial and equatorial profiles do grow -- in fact, these deviations appear to cause an early departure from self-similar solution, so that the third echo $N=3$ is no longer in what we had identified as Phase 2 (see also Fig.~\ref{fig:phi_central}).   We still observe some resemblance with the Choptuik solution for larger values of $R$, and were able to match these features reasonably well with $\Delta = 2.75$ -- clearly, however, this value should be considered a crude estimate only.   At smaller values of $R$, however, and on the axis, a new feature appears to have emerged by the time of the third echo.  

In Fig.~\ref{fig:echoes_075} we included results for two angular resolutions $N_\theta = 14$ and $N_\theta = 12$.  The two resolutions cannot be distinguished in the figure at early times, as well as at sufficiently large radius.  The new features that emerge late in the evolution, however, appear on the axis, and therefore have a strong angular dependence that is difficult to resolve in spherical coordinates.  While this leads to increasing quantitative differences between the different angular resolutions, the qualitative behavior remains very similar, as we discuss in more detail in the following Section.  (The ``double peak" peak at $e^{3 \Delta} R \simeq 10$ for $N=3$, $\theta = 0$ and $N_\theta = 12$ leads to a single peak very similar to that for $N_\theta = 14$, just at a slightly later time; compare Fig.~\ref{fig:bifurcation}.)

\subsubsection{Properties of the deformations}
\label{sec:deformations}

\begin{figure}[t]
\begin{center}
\includegraphics[width=3.5in]{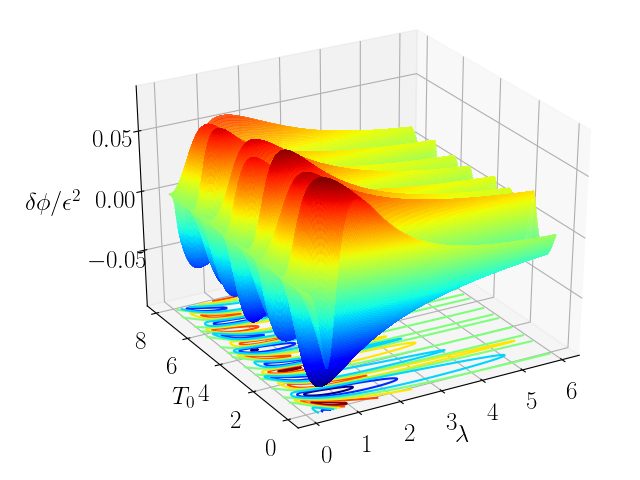}
\end{center}
\caption{Differences in the scalar field $\phi$ as observed by pairs of photons emitted from the origin at logarithmic times $T_0$ in the axial and equatorial directions, as a function of the photons' affine parameter $\lambda$ (see Eq.~\ref{delta_phi}).  Shown are results for a near-critical evolution with $\epsilon^2 = 0.01$.}
\label{fig:delta_phi}
\end{figure}

\begin{figure}[t]
\begin{center}
\includegraphics[width=3in]{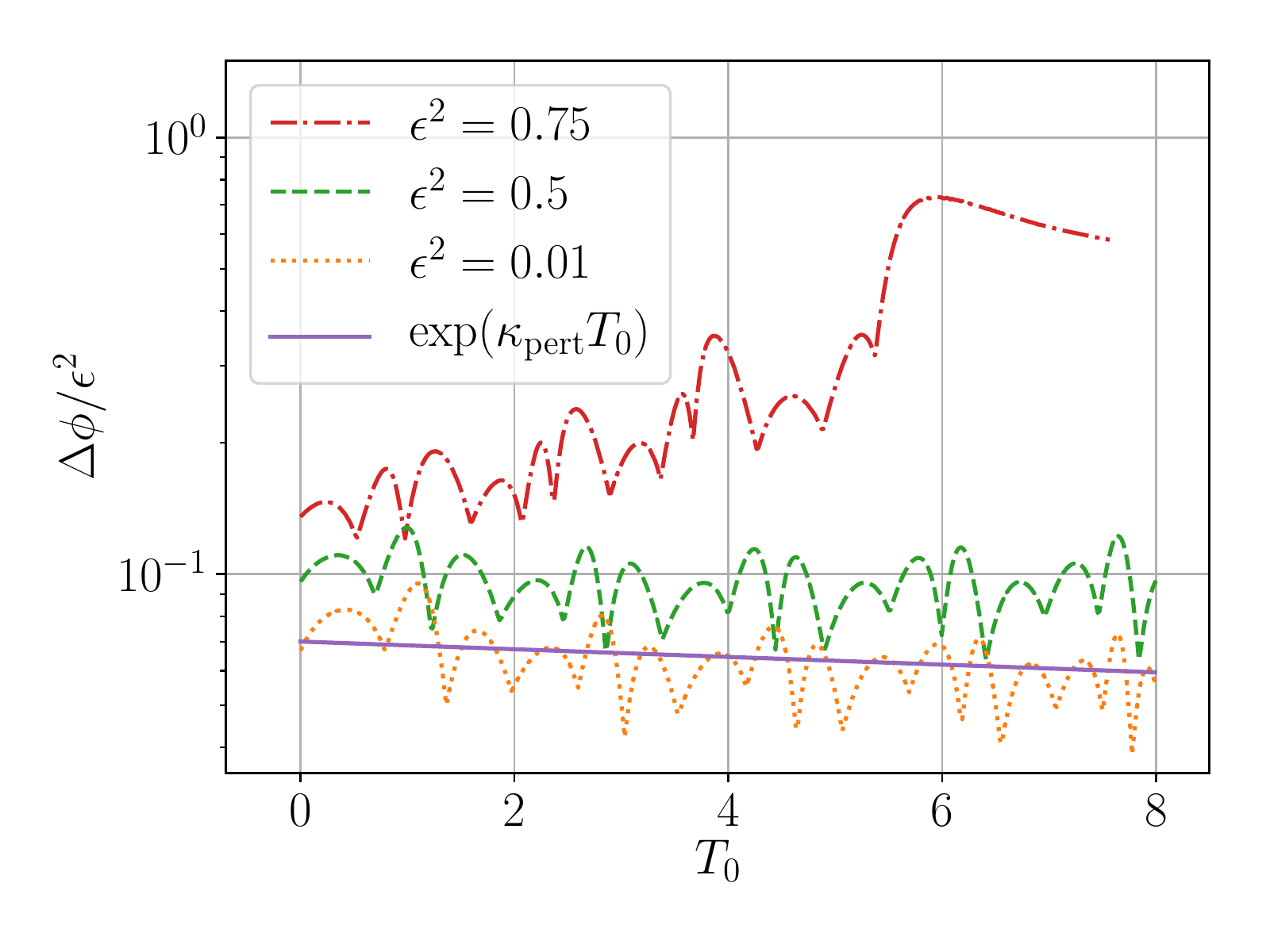}
\end{center}
\caption{Maximum differences in the scalar field $\phi$ as observed by pairs of photons emitted from the origin at times $T_0$ in the axial and equatorial direction (see Eq.~\ref{Delta_phi}), for near-critical evolutions with different values of $\epsilon^2$.  The solid line shows the decay rate for even $\ell = 2$ modes as predicted by MGG.}
\label{fig:Delta_phi}
\end{figure}

In order to analyze the growth or decay of deviations from spherical symmetry more carefully we emit pairs of photons from the origin in the axial and equatorial directions, as described in Section \ref{sec:diagnose}.  We then record the differences $\delta \phi$ in the scalar field $\phi$ that these pairs observe as a function of the photons' affine parameter $\lambda$ (see Eq.~\ref{delta_phi}), and also compute the maximum values of the differences $\Delta \phi$ (see Eq.~\ref{Delta_phi}).   In Fig.~\ref{fig:delta_phi} we show results for $\delta \phi$ for a near-critical evolution with $\epsilon^2 = 0.01$, and in Fig.~\ref{fig:Delta_phi} maximum values $\Delta \phi$ for different values of $\epsilon$ (rescaled with $\epsilon^2$).

For spherical evolutions, with $\epsilon = 0$, the differences $\Delta \phi$ always remain close to truncation error (about $5 \times 10^{-14}$ or less).  This behavior is different from that reported by CHLP, who find a growth in $\Delta \phi$ even for $\epsilon = 0$, although these values of $\Delta \phi$ decrease with improving numerical accuracy (see their Fig.~6).  This effect might be related to a small drift of the center of symmetry with respect to the origin in the simulations of CHLP, which also converges to zero as numerical accuracy is improved.   We do not observe such a drift in our simulations with spherical polar coordinates and equatorial symmetry, nor a growth in $\Delta \phi$ for $\epsilon = 0$.  

According to MGG, linear perturbations of the Choptuik spacetime can be described by functions of the form
\begin{equation} \label{perturb}
u(T) \simeq e^{\kappa T} \cos(\omega_p T + \phi_{\rm ph}) g(T),
\end{equation}
where $\kappa$ determines the rate of perturbation's growth or decay.  The function $g(T)$ is a function with the period $\Delta$ of the discretely self-similar background solution.   Values of $\kappa$ and $\omega_p$ for different modes $\ell$ are tabulated in Table~I of MGG.  In particular, MGG found that all $\kappa$ are negative, leading them to conclude that ``all nonspherical perturbations of the Choptuik spacetime decay."  

In general, the period $2 \pi / \omega_p$ of the perturbation is not commensurate with the period $\Delta$ of the background, making it difficult to analyze theses functions, especially if data are available for only a small number of periods $\Delta$ (see also the discussion in MGG).   This situation is quite different from the critical collapse of ultrarelativistic fluids, for which the critical solution is continuously self-similar, so that that perturbations behave like (\ref{perturb}) but without the function $g(T)$ (see \cite{Gun02}).  In this case the perturbations behave like damped or growing oscillations, and the coefficients $\kappa$ and $\omega_p$ can be determined from fits to the numerical data (see \cite{BauM15,CelB18}). 

While it is significantly harder to determine $\kappa$ and $\omega_p$ for scalar fields, the curve for $\epsilon^2 = 0.01$ in Fig.~\ref{fig:Delta_phi} does appear like a slowly damped oscillation (the same damping is present in Fig.~\ref{fig:delta_phi}, but more difficult to see in a surface plot).  For comparison, we included the exponential $\exp(\kappa_{\rm pert} T)$ with $\kappa_{\rm pert}  = - 0.07 / \Delta$, the value for an even $\ell = 2$ mode according to MGG (see their Table~I).   The slope of this curve appears to agree quite well with the overall decay of our numerical curve for $\Delta \phi$.  

\begin{figure}[t]
\begin{center}
\includegraphics[width=3in]{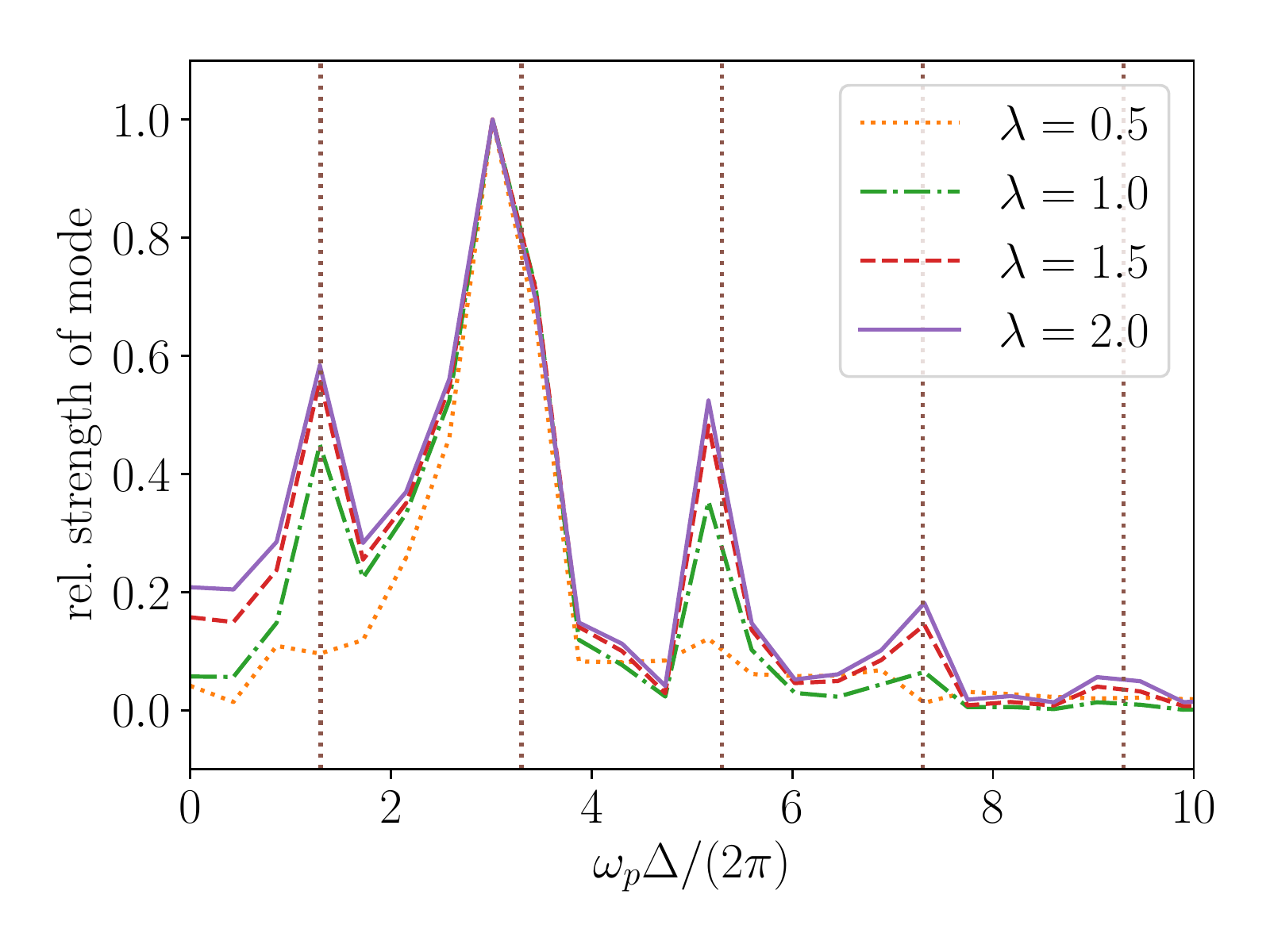}
\end{center}
\caption{The low-frequency end of the discrete Fourier transform of $\exp(- \kappa_{\rm pert} T_0) \, \delta \phi$ for several different fixed values of $\lambda$, for a near-critical evolution with $\epsilon^2 = 0.01$ (i.e.~the Fourier transform along lines of constant $\lambda$ in Fig.~\ref{fig:delta_phi}, except that we removed the overall exponential decay $\exp(\kappa_{\rm pert} T_0)$ from $\delta \phi$).  The vertical dotted lines at frequencies $\omega_p \Delta / (2 \pi) = 1.3$, 3.3, 5.3 \ldots mark the peaks in the spectra as expected from the perturbative calculations of MGG for even $\ell = 2$ modes of the scalar field (see their Fig.~2).}
\label{fig:spectrum}
\end{figure}

Following MGG, we now multiply $\delta \phi$ for $\epsilon^2 = 0.01$ with $\exp(- \kappa_{\rm pert} T_0)$ in order to remove the overall exponential decay, and then take the Fourier transform of $\exp(- \kappa_{\rm pert} T_0) \, \delta \phi$ for fixed values of $\lambda > 0$ ($\lambda = 0$ corresponds to the origin, where $\delta \phi = 0$ identically).  The low-frequency end of these Fourier transforms is shown in Fig.~\ref{fig:spectrum}; for larger frequencies the spectra quickly decay to zero.  Given that we have data for $T_0 \lesssim 3 \Delta$ only, the resolution of these spectra is quite crude.   We can nevertheless observe well-defined peaks, and moreover these peaks agree well with those predicted in the perturbative calculations of MGG, marked by the vertical dotted lines in Fig.~\ref{fig:spectrum}.   We note that the value of $\omega_0 \Delta / (2 \pi) = 0.3$ provided in their Table I is by definition in the range between 0 and 2 (i.e.~it is defined modulo two).  Peaks in the spectrum of the scalar field perturbations can then be found by adding odd multiples of $2 \pi / \Delta$ to $\omega_0$, yielding values of $\omega_p \Delta / (2 \pi) = 1.3$, 3.3, 5.3 \ldots (compare their Fig.~2), while peaks in the spectrum of the metric perturbation can be found by adding even multiples of $2 \pi / \Delta$  (see also the discussion around their Eq.~173).  We conclude that, for small deformations $\epsilon^2$, our results are consistent with both the slow exponential decay and the oscillation rates of MGG.

For larger deformations $\epsilon^2$, however, we find deviations.  For $\epsilon^2 = 0.5$, the curve in Fig.~\ref{fig:Delta_phi} does not seem to decay at all, which is qualitatively consistent with our observations from Fig.~\ref{fig:echoes_05}.  For $\epsilon^2 = 0.75$ the deviations grow, consistent with Fig.~\ref{fig:echoes_075}, which suggests that $\kappa$ has changed sign and is now positive.

It may seem surprising that $\kappa$ depends on the size $\epsilon$ of deviations from sphericity, since it describes linear perturbations of the unperturbed, spherically symmetric background solution, the Choptuik spacetime, and hence reflects properties of the latter.  However, we have already confirmed the observation of CHLP that other coefficients describing the properties of the background solution, namely $\gamma$ and $\Delta$, also appear to depend on $\epsilon$.  In this sense the $\epsilon$-dependence of $\kappa$ is consistent with that of the others.  Unlike the others, the decay constant $\kappa$ changes sign, but that may be a result of it starting out very close to zero.   We also note that a similar effect was observed for rotating ultrarelativistic fluids, where the decay rate of $\ell = 1$ modes also appeared to show some dependence on the rotation rate (see \cite{GunB18}).   Strictly speaking, the coefficients $\kappa$, $\gamma$ and $\Delta$ should be defined for small deviations from spherical symmetry only, so that the evolution can be approximated as the Choptuik spacetime plus a linear perturbation.  The results of CHLP, \cite{GunB18}, and ours here suggest that for larger values of the deviation from sphericity the evolution can, heuristically, still be described in terms of similar parameters, except that nonlinear terms in the deviation lead to changes in the effective values of these coefficients.  

\begin{figure}[t]
\begin{center}
\includegraphics[width=3in]{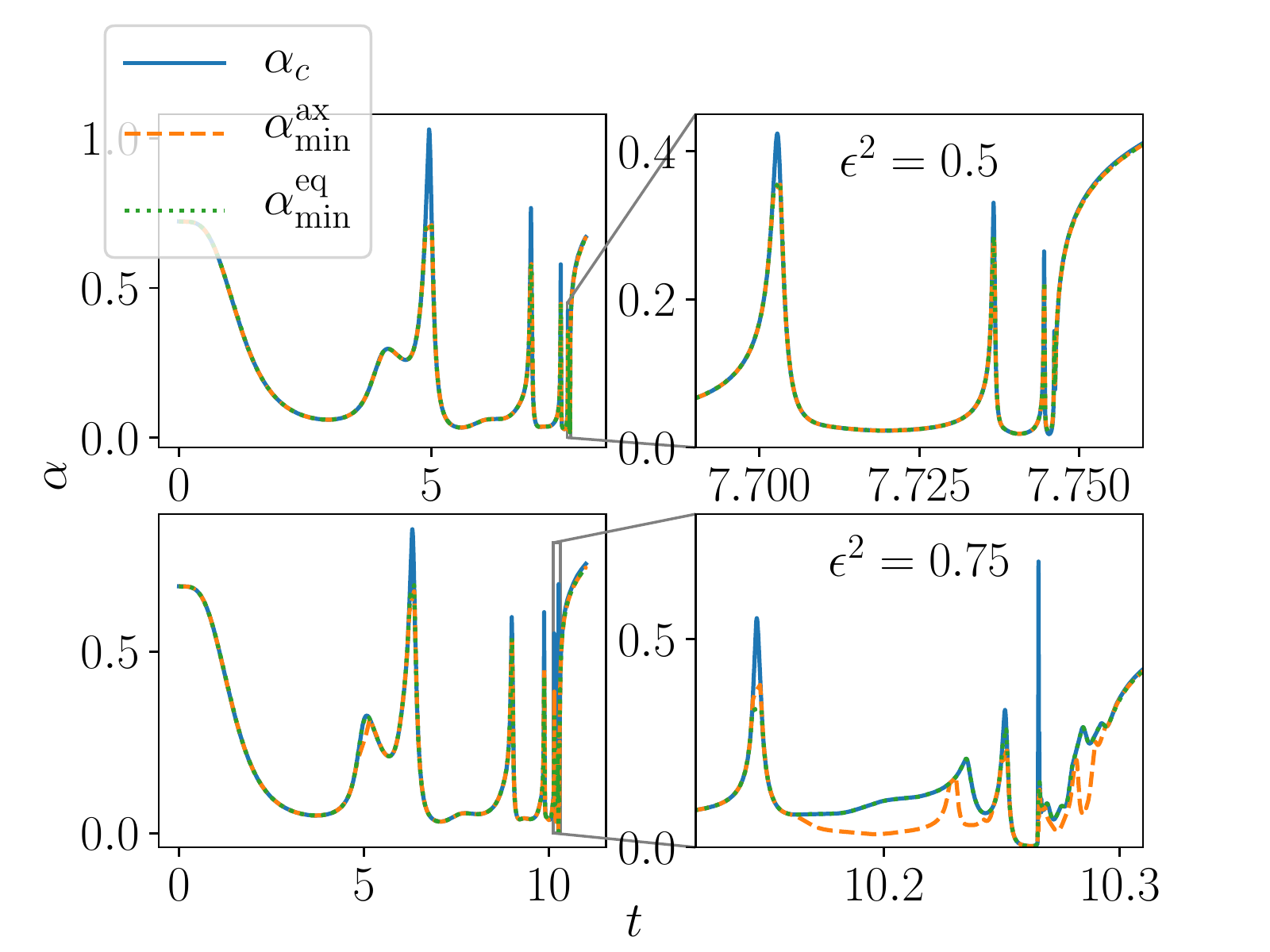}
\end{center}
\caption{The lapse function $\alpha$ as a function of coordinate time $t$ for near-, but subcritical evolutions with $\epsilon^2 = 0.5$ (top panels) and $\epsilon^2 = 0.75$ (bottom panels).  The (blue) solid lines represent the value of the lapse at the center, similar to the solid blue line in Fig.~\ref{fig:lapse_spherical} for $\epsilon^2 = 0$.  The (red) dashed lines mark the minimum values of the lapse along the axis ($\theta = 0$), while the (green) dotted lines mark the minimum values in the equatorial plane ($\theta = \pi/2$).}
\label{fig:lapse}
\end{figure}

We next present some evidence that the growing modes lead to a bifurcation in the solution, similar to that described by CHLP.  A first hint is offered by the lapse function $\alpha$.  In Fig.~\ref{fig:lapse} we show graphs of the lapse function for subcritical evolutions close to criticality, both for $\epsilon^2 = 0.5$ (top panels) and $\epsilon^2 = 0.75$ (bottom panels).  In all panels we show the value of the lapse at the center as a function of coordinate time ($\alpha_c$, blue solid lines), as well as the minimum values of the lapse along the axis ($\alpha_{\rm min}^{\rm ax}$, red dashed lines) and in the equatorial plane ($\alpha_{\rm min}^{\rm eq}$, green dotted lines).  For $\epsilon^2 = 0.5$, the graph is qualitatively very similar to that in Fig.~\ref{fig:lapse_spherical} for spherical evolutions.  In particular we see that all three curves coincide at times when the lapse takes a minimum, indicating that the lapse takes its smallest values at the origin.   For $\epsilon^2 = 0.75$, on the other hand, the behavior is qualitatively different.   The evolution no longer shows the same pattern repeating at increasingly small scales; instead the pattern changes at late times.  Moreover, we now see periods of time when the minimum value of the lapse along the axis no longer coincides with its value at the center.  This indicates that the lapse takes a smaller value somewhere on the symmetry axis, suggesting that a new collapsing region emerges away from the center.

\begin{figure}[t]
\begin{center}
\includegraphics[width=3in]{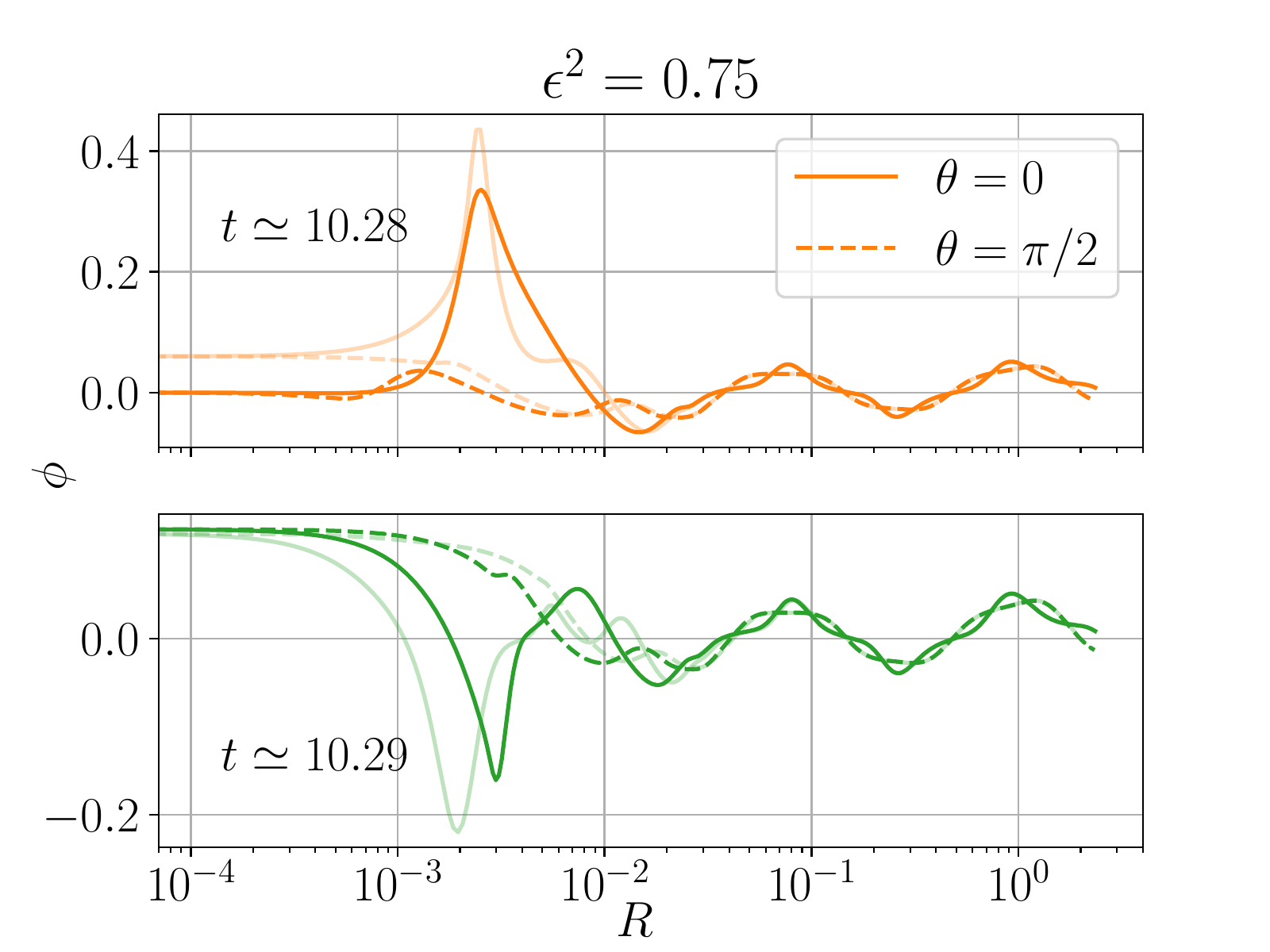}
\end{center}
\caption{Radial profiles of the scalar field $\phi$ for near-critical evolutions with $\epsilon^2 = 0.75$ at times that correspond to the appearance of the first maximum and the first minimum of $\phi$ after the $N=3$ echo.  We show results for both $N_\theta = 14$ (solid colors) and $N_\theta = 12$ (faded colors).  For $N_\theta = 14$, the two coordinate times (proper times at the origin) are $t = 10.2815$ ($\tau = 2.07515$) and $t = 10.2891$ ($\tau = 2.07713$), while for $N_\theta = 12$ they are 10.2865 (2.07716) and 10.2959 (2.07932).}
\label{fig:bifurcation}
\end{figure}

\begin{figure}[t]
\begin{center}
\includegraphics[width=3.5in]{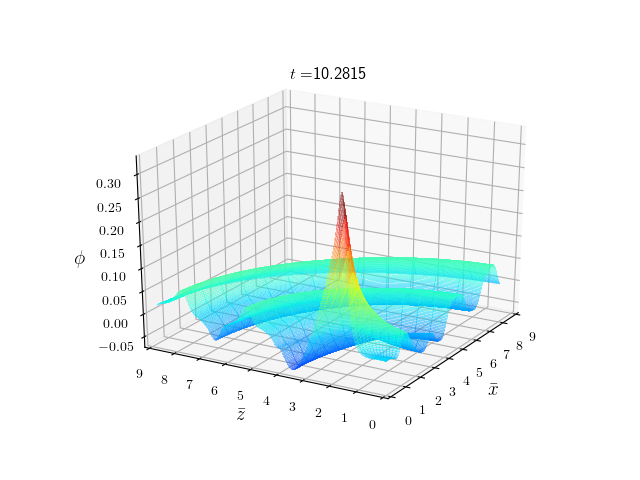}
\includegraphics[width=3.5in]{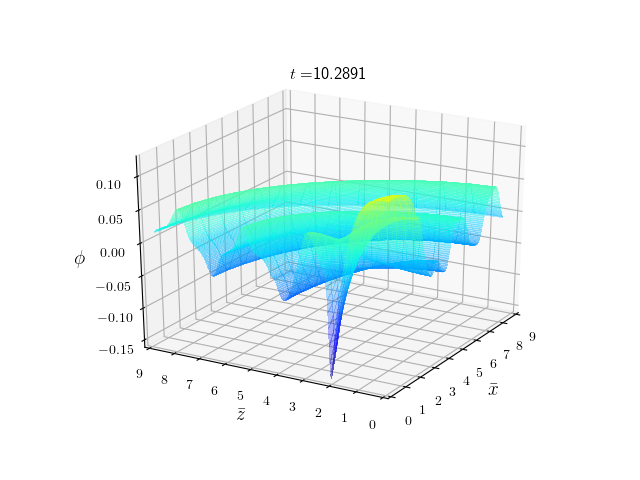}
\end{center}
\caption{Same as Fig.~\ref{fig:bifurcation}, but as surface plots.  Following CHLP we have defined $\bar x = \bar R \sin \theta$ and $\bar z = \bar R \cos \theta$ with $\bar R = \ln(1 + R/e_0)$ and $e_0 = 2 \times 10^{-4}$.  (Compare Fig.~2 in CHLP.)}
\label{fig:snapshots}
\end{figure}

This is consistent with the behavior of the scalar field $\phi$, which, at late times, starts to develop large oscillations at a point on the symmetry axis.  The example shown in Fig.~\ref{fig:bifurcation} shows such an oscillation at times shortly after the $N= 3$ panel in Fig.~\ref{fig:echoes_075} where, as we discussed before, a similar feature can be seen.  In Fig.~\ref{fig:snapshots} we show $\phi$ at the same times as surface plots.

Recall that we imposed equatorial symmetry in our simulations.  This means that if a feature develops at a certain distance from the origin on the symmetry axis ($\theta = 0$ in Fig.~\ref{fig:bifurcation}, or the $\bar z$-axis in Fig.~\ref{fig:snapshots}), then the same feature also develops at the same distance on the other side of the equatorial plane.  Our simulations therefore suggest the formation of two centers of collapse on the symmetry axis, one on each side of the origin.  This behavior was first observed by CHLP, who referred to this process as a bifurcation.

The plots in Fig.~\ref{fig:snapshots} also demonstrate, however, that our numerical grid is not sufficient to accurately resolve a small feature developing away from the origin -- exactly as CHLP had anticipated: ``Of course, spherical polar coordinates would not be well suited to following a solution beyond a bifurcation, but it should be adequate to study the growth or decay of perturbations."  We showed most results for $\epsilon^2 = 0.75$ with $N_\theta = 14$, but also compared with results for $N_\theta = 12$, the resolution used in our other aspherical simulations.  In Fig.~\ref{fig:echoes_075}, for example, the two resolutions can hardly be distinguished except in the presence of the new feature that grows on the axis, and which leads to an increasing angular dependence.  We similarly include results for both angular resolutions in Fig.~\ref{fig:bifurcation}.  Note that these simulations differ not only in angular resolution, but also in precise distance from the critical parameter $\eta_*$, which makes a direct comparison difficult.  Despite the evident quantitative differences, we find very similar qualitative behavior, namely the formation of new centers of oscillation on the symmetry axis away from the origin.  While our numerical resolution is not sufficient to analyze these features in greater detail, our simulations suggest that, for large initial deviations from sphericity, perturbations of the Choptuik spacetime can grow, and that these growing perturbations may lead to a bifurcation in the solution as previously observed by CHLP.


%
%

\section{Summary}
\label{sec:summary}

We perform numerical simulations of the critical collapse of massless scalar fields in the absence of spherical symmetry.   We use a numerical relativity code that adopts spherical coordinates with a logarithmic radial coordinate and, following CHLP, evolve axisymmetric initial data whose deviation from sphericity is parameterized by $\epsilon$.  

For small $\epsilon$ we find values for the critical exponent $\gamma$ and the echoing period $\Delta$ that agree well with established values.  We also find that small deviations from spherical symmetry behave like slowly damped oscillations, with a decay rate $\kappa$ and frequencies $\omega_p$ similar to those found by MGG in perturbative calculations.   

For larger deviations from spherical symmetry we find effective values for the critical exponent $\gamma$ and the echoing period $\Delta$ that decrease with increasing $\epsilon$, in good agreement with the findings of CHLP.    Moreover, we find that, for sufficiently large departures from spherical symmetry, the deviations grow, rather than decay, also confirming earlier results of CHLP.  This suggests that the effective value for the decay rate $\kappa$, like those for $\gamma$ and $\Delta$, depends on $\epsilon$ also, and, in fact, changes sign.  We find some evidence that, at late times, these growing modes lead to a bifurcation of the critical solution, with large oscillations of the scalar field developing around two points on the symmetry axis, one above and one below the origin.   While our numerical grid does not provide sufficient resolution away from the center to follow the evolution of these oscillations, and possibly their collapse to black holes (unlike the code of CHLP, our code does not employ AMR), our simulations nevertheless display the growth of these features, and hence support the original discovery of this bifurcation by CHLP.

A similar phenomenon was recently discussed by \cite{HilWB17} in the context of collapse of axisymmetric gravitational waves (see also \cite{AbrE93,Sor11}).  Specifically, \cite{HilWB17} evolved Brill wave initial data and  found that, close to criticality, the Kretschmann scalar takes maxima on the symmetry axis away from the origin, reminiscent of the results of CHLP as well as ours here.   Moreover, \cite{HilWB17} find that, in their equatorially symmetric simulations, two black holes with disjoint apparent horizons form on the axis, suggesting that near-critical Brill wave initial data lead to the formation of two black holes which then collide head-on.   The authors further suggest that this behavior may dominate whenever axisymmetric modes dominate over spherically symmetric modes.

Unlike data for gravitational waves, the initial data (\ref{indata}) can be produced as an expansion in $\epsilon$ about spherically symmetric data.  To linear order, one would therefore expect agreement with the perturbative results of MGG, and indeed, for small $\epsilon$ we observe a decay rate $\kappa$ and oscillation frequencies $\omega_p$ that are similar to the values computed by MGG.  For larger $\epsilon$, however, we observe shifts in the effective values of the critical exponent $\gamma$, the echoing period $\Delta$ as well as the decay rate $\kappa$.  Similar behavior was also observed by \cite{GunB18} for rotating ultrarelativistic fluids, even though there the effect of increasing rotation was to stabilize the perturbations, whereas here the effect of increasing deformations is to destabilize the perturbation \footnote{There exists a well-known equivalence between irrotational stiff fluids and scalar fields, which might mislead one to conclude that critical phenomena in one system ought to behave exactly as those in the other.  This is not the case, however; in fact, as clarified by \cite{BraCGN02}, the critical solution in one system is not observed in the other, and vice versa.}.  The coefficients $\gamma$ and $\kappa$ have a well-defined meaning only in the context of linear perturbations.  Heuristically, however, it appears that a similar description of the dynamics is still possible in the presence of nonlinear effects, except that the effective coefficients $\gamma$ and $\kappa$ then depend on the size of the perturbation.  Exploring this possibility quantitatively would require (at least) a second-order perturbation analysis, which, unfortunately, is quite challenging. 


\acknowledgments

It is a great pleasure to thank Carsten Gundlach for numerous elucidating conversations and for suggesting the spectral analysis in Fig.~\ref{fig:spectrum}.  I would also like to thank him, David Hilditch, Steve Liebling and Frans Pretorius for detailed and very helpful comments on earlier versions of this manuscript.   This work was supported in part by NSF grants PHY-1402780 and 1707526 to Bowdoin College, and through sabbatical support from the Simons Foundation (Grant No.~561147 to TWB).  Numerical simulations were performed on the Bowdoin Computational Grid.

\end{document}